\documentclass[a4paper,11pt]{article}
\pdfoutput=1 

\usepackage{jheppub} 
\usepackage[table]{xcolor}
\usepackage[english]{babel}
\usepackage[utf8]{inputenc}
\usepackage{amsmath}
\usepackage{graphicx}
\usepackage{amsmath}
\usepackage{indentfirst}
\usepackage{color}
\usepackage{amssymb}
\usepackage{array} 
\usepackage{float}
\newcolumntype{M}[1]{>{\centering\arraybackslash}m{#1}}
\newcolumntype{N}{@{}m{0pt}@{}}

\def\be{\begin{equation}}
\def\ee{\end{equation}}
\def\bea{\begin{eqnarray}}
\def\eea{\end{eqnarray}}

\newcommand{\mc}{\mathcal}

\title{On the choice of entropy variables in multifield inflation}

\author[a,b]{Michele Cicoli,}
\author[c]{Veronica Guidetti,}
\author[d]{Francesco Muia,}
\author[a,b]{Francisco G. Pedro,}
\author[b]{Gian Paolo Vacca}
\affiliation[a]{\footnotesize Dipartimento di Fisica e Astronomia, Universitá di Bologna, via Irnerio 46, 40126 Bologna, Italy}
\affiliation[b]{\footnotesize INFN, Sezione di Bologna, via Irnerio 46, 40126 Bologna, Italy}
\affiliation[c]{\footnotesize Deutsches Elektronen-Synchrotron (DESY), 22607 Hamburg, Germany}
\affiliation[d]{\footnotesize DAMTP, University of Cambridge, Wilberforce Road, Cambridge, CB3 0WA, UK}

\emailAdd{michele.cicoli@unibo.it}
\emailAdd{veronica.guidetti@desy.de}
\emailAdd{fm538@damtp.cam.ac.uk}
\emailAdd{francisco.soares@unibo.it}
\emailAdd{vacca@bo.infn.it}

\abstract{We discuss the usefulness and theoretical consistency of different entropy variables used in the literature to describe isocurvature perturbations in multifield inflationary models with a generic curved field space. We clarify which is the proper entropy variable to be used to match the evolution of isocurvature modes during inflation to the one after the reheating epoch in order to compare with observational constraints. In particular, we find that commonly used variables, as the relative entropy perturbation or the one associated to the decomposition in tangent and normal perturbations with respect to the inflationary trajectory, even if more useful to perform numerical studies, can lead to results which are wrong by several orders of magnitude, or even to apparent destabilisation effects which are unphysical for cases with light kinetically coupled spectator fields.}

\begin{document} 
\hfill DESY 21-106

\maketitle
\flushbottom

\section{Introduction}
\label{Intro}

The development of cosmic inflation~\cite{Guth:1980zm, Linde:1981mu, Albrecht:1982wi} to solve some of the issues of the standard Big Bang theory and as a mechanism to generate the primordial perturbations observed in the Cosmic Microwave Background (CMB) has been one of the most striking results of cosmology during the last decades \cite{Baumann:2009ds, Akrami:2018odb}. In its simplest implementation, i.e. in single field models, inflation can in principle account for all observations. Specific single field models have been increasingly constrained over the years: some of the simplest models such as power-law inflation~\cite{Linde:1983gd} are now strongly disfavoured and - in general - models featuring concave potentials are favoured with respect to those with convex potentials~\cite{Akrami:2018odb}. 

However, it is reasonable to expect that the single field description of inflation is valid only approximately as in principle there may be many fields that couple to the inflaton and be dynamical during inflation. For instance, UV complete theories like string theory give rise, upon compactification, to hundreds of complex scalar fields in the low-energy 4D theory whose dynamics can be crucial during inflation. Moreover, as typical of supergravity models, the resulting field space is curved. Notice that, in a gravitational context, theories with a curved field space and non-trivial kinetic interactions are not just a feature of string theory but generically arise in non-minimally coupled multifield effective models, upon moving to Einstein frame~\cite{Kaiser:2010ps,Kaiser:2012ak, Schutz:2013fua, DiMarco:2002eb}. 

Some of these scalars might acquire a very large mass and decouple completely from the inflationary dynamics. However several of them are expected to remain light and to play an important role during inflation. A primary example are axion-like fields which are naturally light since they enjoy a shift symmetry which is broken only at non-perturbative level. Again, a typical string compactification is characterised by hundreds of them, leading to the so-called string axiverse~\cite{Svrcek:2006yi, Conlon:2006tq, Arvanitaki:2009fg, Cicoli:2012sz}. Therefore in string models, and more generally in non-minimally coupled effective theories, it is very natural to have light kinetically coupled fields during inflation, that in principle cannot be decoupled from the inflationary dynamics. It becomes then crucial to study the effect of these isocurvature modes on the inflationary dynamics, to understand if they can modify or even spoil inflation, depending on their observational signatures.

The choice of entropy variables to describe the effect of isocurvature modes in multifield models of inflation is clearly fundamental. Several variables have been proposed so far in the literature, with different pros and cons. In this paper we will discuss the usefulness and theoretical consistency of each of these entropy variables, clarifying which is the one more suitable to perform a proper matching between the inflationary and the post-inflationary evolution of isocurvature perturbations. This matching is crucial to make contact with data since CMB observations place constraints on the primordial isocurvature fraction only at the time of CMB decoupling. Let us summarise our analysis:
\begin{itemize}
\item Planck measurements constrain the primordial isocurvature fraction parameter defined as $\beta_{\rm iso}(k) = P_{S_{n\gamma}}(k)/ [P_\zeta(k) + P_{S_{n\gamma}}(k)]$, where $\zeta$ is the standard curvature perturbation on uniform-density hypersurfaces, while $S_{n\gamma}$ is the relative entropy perturbation between photons and a different species denoted with $n$ which can be cold dark matter, baryons or neutrinos. The upper bound on this fraction goes from about $1\%$ to $10\%$ depending on the species involves and the momentum scale \cite{Akrami:2018odb}.

\item The correct entropy variable to be studied during inflation seems therefore $S_{ij}$, where now $i$ and $j$ are indices denoting different fields (or combinations thereof) of the underlying multifield models. The evolution of $S_{ij}$ during inflation should be matched with the post-inflationary history of our universe, leading to $S_{n\gamma}$, once a proper understanding of the model-dependent reheating epoch is developed \cite{Polarski:1994rz, Langlois:1999dw}. This involves the knowledge of how photons, cold dark matter, baryons and neutrinos are produced from the decay of the different microscopic degrees of freedom. This study has been recently carried out in flat field spaces \cite{Martin:2021frd} but the extension of this formalism to curved systems is still missing. 

\item There are some particular cases where however the relative entropy variable $S_{ij}$ becomes ill-defined since it would apparently diverge. This might occur in general in multifield models where the kinetic energy of some scalar fields vanishes \cite{Wands:2000dp} as in situations with a curved field space and light spectator fields \cite{Cicoli:2018ccr, Cicoli:2019ulk} since $S_{ij}\sim \left(\frac{\delta\rho_j}{\dot\rho_j} - \frac{\delta\rho_i}{\dot\rho_i}\right) \to \infty$ due to $\dot\rho_i\to 0$ (where $\rho_i$ is the energy density of the $i$-th scalar field, $\delta\rho_i$ is its perturbation and $\dot\rho_i \equiv d\rho_i/dt$).

\item We will show that this divergence is unphysical since no geometrical destabilisation \cite{Renaux-Petel:2015mga} of the background trajectory is induced by the spurious growing isocurvature perturbations. In fact, we will argue that in this case a better entropy variable which should be used is $\hat{S}_{ij}\propto \dot\rho_i\dot\rho_j S_{ij}$ which remains well-behaved and decreases with respect to standard adiabatic perturbations. In these `pathological' cases it is therefore the evolution of $\hat{S}_{ij}$, and not the one of $S_{ij}$, which should be studied during inflation. Matching with the beginning of radiation dominance will then lead to $\hat{S}_{n\gamma}$ which can be safely translated into $S_{n\gamma}$ to compare with observational data since $S_{n\gamma}$ is a well-defined quantity with no divergence.

\item Another entropy variable which is sometimes used is the total entropy defined in terms of the non-adiabatic pressure perturbation $\delta P_{\rm nad}$ as $S = \left(H/\dot P\right)\delta P_{\rm nad}$. This quantity, even if used in some cases \cite{Huston:2011fr, Huston:2013kgl}, contains both intrinsic and relative entropy contributions, and so it is not the right one to be used to compare with observational bounds on $\beta_{\rm iso}$.

\item The non-adiabatic perturbation $\delta s$, introduced for the first time in \cite{Gordon:2000hv} and then used extensively in many studies (see for example \cite{Achucarro:2010da, Cremonini:2010sv}), corresponds instead to fluctuations which are orthogonal to the classical background trajectory. The entropy variable $\delta s$ is very useful to picture clearly the decomposition of an arbitrary perturbation into adiabatic (tangent) and entropy (normal) components, and it allows for an easier numerical integration of the field equations. However it is not the right variable to compare with observations since it can lead to results which are wrong by several orders of magnitude since $\delta s$ is proportional to total entropy \cite{Huston:2011fr, Huston:2013kgl, Wands:2002bn}. Moreover, in a curved field space, the use of $\delta s$ can lead to apparent destabilisation effects which are unphysical. This effect can be seen in the behaviour of the effective mass-squared $m_{\rm eff}^2$ of isocurvature perturbations $\delta s$ which receives different contributions from derivatives of the scalar potential, the curvature of the field space and the turning rate of the background trajectory. Depending on the form of the kinetic coupling and the underlying scalar potential, there are 3 general situations: 
\begin{enumerate}
\item Multifield models featuring a geometrical destabilisation \cite{Renaux-Petel:2015mga} induced by the negative curvature of the field space that causes a tachyonic effective mass \cite{Gong:2011uw} which however turns into $m_{\rm eff}^2>0$ once the system settles down in the classical background trajectory \cite{Cicoli:2018ccr, Christodoulidis:2019mkj}. In this case both $S_{ij}$ and $\delta s$ would decay since the isocurvature modes are heavy (even if $S_{ij}$ is the right variable to study to get to the final prediction for $\beta_{\rm iso}$ after reheating).

\item Multifield models with a non-vanishing turning rate \cite{Achucarro:2016fby} characterised by $m_{\rm eff}^2\simeq 0$ due to cancellation between different contributions even if the background field can be heavy (i.e. the eigenvalues of the Hessian of the scalar potential are non-zero). In this case isocurvature modes are constant on super-horizon scales and act as a source for curvature perturbations. Thus $\beta_{\rm iso}$ is expected to be negligible in this class of models which can qualitatively be well-described both in terms of $S_{ij}$ and $\delta s$, even if one should focus on $S_{ij}$ to pin down the exact prediction for $\beta_{\rm iso}$.

\item Multifield models typical of string compactifications where the inflaton is a K\"ahler modulus \cite{Conlon:2005jm, Cicoli:2008gp, Burgess:2010bz, Cicoli:2011ct, Cicoli:2012cy, Cicoli:2016chb, Cicoli:2016xae, Cicoli:2017axo, Cicoli:2020bao} and ultra-light axions play the role of spectator fields. In the case of Fibre Inflation \cite{Cicoli:2008gp}, as shown in \cite{Cicoli:2019ulk}, when using the entropy variable $\delta s$, the effective mass of these isocurvature modes would become tachyonic due to the negative contribution coming from the curvature of the underlying field space, i.e. $m_{\rm eff}^2<0$. This might signal an exponential growth of isocurvature modes leading the system to a regime where perturbation theory breaks down. However the background trajectory of these multifield models turns out to be fully stable \cite{Cicoli:2019ulk}. The absence of any instability seems therefore to be a paradox. This is resolved by noticing that the variable $\delta s$ is ill-defined in this case since the divergence arises from the behaviour of the vector normal to the background trajectory which is used to decompose an arbitrary perturbation to define $\delta s$. In fact, when using the relative entropy variable $\hat{S}_{ij}$, we will find that in this case non-adiabatic perturbations decay on super-horizon scales, in full agreement with the absence of any destabilisation of the background trajectory. 
\end{enumerate}

\item As we will show for the general case of a curved field space, the different entropy variables, $S$, $S_{ij}$, $\hat{S}_{ij}$ and $\delta s$, are all gauge invariant. Thus the apparent instability for cases with light kinetically coupled spectator fields where both $S_{ij}$ and $\delta s$ diverge, is not due to the use of a non-physical variable which is not gauge invariant, but it is simply due to the fact that in these particular cases those variables become ill-defined.
\end{itemize} 

This paper is organised as follows. In Sec.~\ref{sec:EntropyPerturbations} we discuss different variables used in the literature to describe entropy perturbations in multifield inflationary models, stressing which is the right variable to consider to perform a proper matching between inflation and the reheating epoch. In Sec. \ref{sec:SpuriousInstability} we show how some entropy variables can become ill-defined in models with kinetically coupled spectator fields which are lighter than the inflaton. Focusing on a simple 2-field system with a Starobinsky-like inflationary potential, we investigate the origin of the spurious instability and we then show that, using the appropriate entropy variable, the isocurvature power spectrum turns out to be negligible. We draw our conclusions in Sec.~\ref{sec:Conclusions} and we leave some technical details to the appendices. In particular, in App. \ref{AppA} we first present the general framework of perturbation theory for a curved field space (focusing on the spatially flat gauge), and we then prove gauge invariance of all entropy variables. App. \ref{sec:singlefield} provides instead the details of the single field limit of density perturbations for our illustrative 2-field model.

\section{Entropy variables in multifield inflation}
\label{sec:EntropyPerturbations}

In this section we clarify the relations between the various definitions of entropy that can be found in the literature, in an effort to understand how the entropy perturbations generated during multifield inflation are then transferred to the primordial plasma in the radiation phase.

\subsection{Entropy perturbations during inflation}
\label{sec:PerturbationsDuringInflation}

Let us start with the multifield Lagrangian of a generic non-linear sigma model which can be written as:
\be
\frac{\mc{L}}{\sqrt{|g|}}=\frac12\, G_{ij}(\phi^i)\partial_\mu \phi^i \partial^\mu \phi^j-V(\phi^i)\,,
\label{eq:L}
\ee
where $G_{ij}(\phi^i)$ denotes the field space metric. As already mentioned, this class of models naturally arises in the framework of beyond the Standard Model theories as string compactifications, supergravity and non-minimally coupled multifield effective theories. 

To make contact with the entropy $S$, let us recall that for a generic fluid the pressure $P$ is a function of $S$ and the energy density $\rho$: $P=P(\rho,S)$. Pressure perturbations can therefore be decomposed into an adiabatic and a non-adiabatic part, according to:
\be
\delta P= \frac{\delta P}{\delta \rho}\Big |_S \delta \rho+ \frac{\delta P}{\delta S}\Big|_\rho \delta S\,.
\ee
The adiabatic pressure perturbation is $\delta P_{\rm ad}=c_s^2\, \delta\rho$, with a speed of sound $c_s^2\equiv \dot{P}/\dot{\rho}$. The non-adiabatic pressure perturbation, denoted $\delta P_{\rm nad}$, is formally given by $\delta P_{\rm nad}=\left(\dot{P}/\dot{S}\right)\delta S$, and is in practice computed by subtracting the adiabatic from the total pressure perturbation: 
\be
\delta P_{\rm nad}=\delta P- \delta P_{\rm ad}\,.
\label{eq:Pnad}
\ee

\subsubsection*{Total entropy}

The total entropy perturbation can be defined in terms of the non-adiabatic pressure perturbation $\delta P_{\rm nad}$ as:
\be
S=\frac{H}{\dot{P}}\,\delta P_{\rm nad}=\frac{H}{\dot{P}}\left(\delta P- c_s^2 \delta \rho\right).
\label{eq:S}
\ee
During inflation, when scalar fields dominate the energy content of the Universe, one may write the total entropy of the system in terms of $\phi^i$ and $\delta\phi^i$ and their derivatives. In the case of multifield inflation in a generic curved field space (\ref{eq:S}) becomes (setting $M_p=1$):
\be 
S=-\frac{H}{3H\dot\phi^i\dot\phi_i+2\dot\phi^i V_i}\,\delta P_{\rm nad}\,,
\label{TotS}
\ee
where in the spatially flat gauge $\delta P_{\rm nad}$ is given by:
\be
\delta P_{\rm nad}=-2 V_i \delta \phi^i + \frac{V_i \dot{\phi}^i \delta\phi_j \dot{\phi}^j}{3H^2} -\frac{2}{3H}\frac{V_\ell \dot{\phi}^\ell}{\dot{\phi}_m \dot{\phi}^m}\left[\dot{\phi}_i \delta\dot{\phi}^i +\frac12 \dot{\phi}^i \dot{\phi}^j \partial_k G_{ij} \delta \phi^k +V_i \delta\phi^i \right].
\ee
In App. \ref{AppA} we have shown that the variable $S$ is gauge invariant not just for canonical fields but also for the case of a curved field space.

\subsubsection*{Intrinsic entropy}

For a given fluid $i$ one can define the intrinsic entropy perturbation $S_{{\rm int},i}$ as \cite{Malik:2004tf}:
\be
S_{{\rm int},i}=\frac{H}{\dot{P}}\left(\delta P_i- c_i^2 \delta \rho_i\right),
\label{eq:Sint}
\ee
so that the total intrinsic entropy perturbation of a system with multiple components is:
\be
S_{\rm int}=\sum_i S_{{\rm int},i}\,.
\label{eq:Srel0}
\ee
Notice that $S_{{\rm int},i}=0$ for fluids for which $P_i=P_i(\rho_i)$ and that $S_{\rm int}$ is also a gauge invariant quantity (see again App. \ref{AppA}).
 
\subsubsection*{Relative entropy}

One can also define the relative entropy perturbation as:
\be
S_{\rm rel}=S-S_{\rm int}\,.
\ee
Noting that $\delta P= \sum_i \delta P_i$ and that $\delta \rho= \sum_i \delta \rho_i$ and making use of \eqref{eq:S}, \eqref{eq:Sint} and \eqref{eq:Srel0} one may write \cite{Malik:2004tf}:
\bea
S_{\rm rel}&=& \frac{H}{\dot{P} \dot{\rho}}\sum_{ij}c^2_i \left(\dot{\rho}_j \delta \rho_i-\dot{\rho}_i \delta \rho_j\right) \nonumber \\
&=&-\frac{1}{6 \dot{\rho}\dot{P}}\sum_{ij} \dot{\rho}_i \dot{\rho}_j \left(c_i^2-c_j^2\right)S_{ij} \,,
\label{eq:Sreldef}
\eea
where in the last line we have introduced the standard definition of entropy perturbation between two fluids:
\be
S_{ij}\equiv\zeta_i-\zeta_j=-3 H \left(\frac{\delta \rho_i}{\dot{\rho}_i}-\frac{\delta \rho_j}{\dot{\rho}_j}\right),
\label{eq:Sab}
\ee
as is commonly found in the context of hot Big Bang cosmology, and have defined the individual fluid speed of sound as $c_i^2=\dot{P}_i/\dot{\rho}_i$. Notice that, since the curvature perturbation on uniform $\rho_i$ hypersurfaces, $\zeta_i=-\Psi-H \delta\rho_i/\dot{\rho}_i$ is a gauge invariant quantity (see App. \ref{AppA} for the definition of $\Psi$, the scalar spatial perturbation to the metric tensor, and a proof of gauge invariance for scalar non-linear sigma models), $S_{ij}$ is automatically gauge invariant, even in the presence of energy transfer between the fluids \cite{Malik:2002jb,Malik:2004tf}.

It is worth pointing out that $S_{ij}$ becomes ill-defined when $\dot\rho_i\to 0$ in multifield models with vanishing kinetic energy for some scalar fields \cite{Wands:2000dp} as it occurs in classes of string inflationary models with ultra-light axions \cite{Cicoli:2018ccr, Cicoli:2019ulk}. In this case we consider an `improved' relative entropy variable $\hat{S}_{ij}$ defined as:
\be
\hat{S}_{ij} \equiv -\frac{1}{6 \dot{\rho}\dot{P}} \dot{\rho}_i \dot{\rho}_j \left(c_i^2-c_j^2\right)S_{ij}\qquad \Rightarrow\qquad S_{\rm rel} = \sum_{ij}\hat{S}_{ij}\,.
\label{hatSij}
\ee
Clearly $\hat{S}_{ij}$ remains finite even if $\dot\rho_i\to 0$. 

\subsubsection*{Entropy field}

In order to introduce an alternative description of adiabatic and entropy perturbations, it is convenient to define the tangent and normal vectors to the inflationary trajectory. Focusing for simplicity on a 2-dimensional field space, these are given in terms of the background velocities as \cite{Achucarro:2010da,Cremonini:2010sv}:
\be
T^i\equiv \frac{\dot\phi^i}{\dot\phi_0}\qquad\text{and}\qquad N^i=\frac{s_N(t)}{\sqrt{G_{jk} D_t T^j D_t T^k}}\, D_t T^i\,,
\label{eq:TN}
\ee
with $\dot\phi_0\equiv\sqrt{G_{ij}\,\dot\phi^i\,\dot\phi^j}$ and $D_t T^i=\dot T^i +\Gamma_{jk}^i T^j\dot\phi^k$, where $\Gamma$ is the field space metric connection and $s_N(t)=\pm1$ accounts for the relative orientation between the normal direction and $\dot T^i$. These two projectors are orthogonal to each other and have unit norm with respect to the scalar product defined with the metric $G_{ij}$. In the context of inflationary physics, adiabatic and entropy perturbations can be characterised in terms of the gauge invariant quantities (see App. \ref{AppA}): 
\be
\delta \sigma= T^i \,\delta \phi_i\qquad \text{and}\qquad\delta s= N^i\, \delta \phi_i\,,
\label{deltas}
\ee
where $\sigma$ is the so-called `adiabatic field', while $s$ is the `entropy field' \cite{Gordon:2000hv}. These can be related to the adiabatic and isocurvature perturbations as (in the spatially flat gauge) \cite{Achucarro:2010da}:
\be
\zeta=-\frac{1}{\sqrt{2\epsilon}}\,\delta \sigma\qquad\text{and}\qquad \tilde{S}=\frac{1}{\sqrt{2\epsilon}}\,\delta s\,,
\label{eq:RSTN}
\ee
where $2\epsilon = \left(\dot\phi_0/H\right)^2$. Notice that the entropy direction $\delta s$ can be directly related to the total entropy on super-horizon scales as \cite{Gordon:2000hv}:
\be
S\simeq - \frac{2 V_N}{3\dot\phi_0^2}\, \delta s = - \frac{2 V_N}{3\dot\phi_0 H}\, \tilde S\,.
\label{Sds}
\ee
Moreover, as shown in App. \ref{AppA}, in a 2-field system with a diagonal field space metric with $G_{ii} \equiv G_i$ for $i=1,2$, $\delta s$ takes the simple expression:
\be
\delta s  = \sqrt{G_1 G_2} \left(\frac{\dot{\phi^1} \dot{\phi^2}}{\dot\phi_0}\right) \delta\phi^{12} \,,
\label{eq:deltas}
\ee
where $\delta\phi^{12}$ is a gauge invariant quantity called `generalised entropy' in \cite{Gordon:2000hv} and defined as:
\be
\delta \phi^{12}\equiv\frac{\delta \phi^1}{\dot{\phi}^1}-\frac{\delta \phi^2}{\dot\phi^2}\,.
\label{dphi12}
\ee
Finding and solving the evolution equation for $\delta s$ not only gives an intuitive picture of the entropy perturbations but is also a more robust manner of numerically computing entropy perturbations than subtracting from the total pressure perturbation its adiabatic component, as in \eqref{eq:Pnad} \cite{Gordon:2000hv, Price:2014xpa}. For these reasons it has become the preferred way of describing entropy perturbations during multifield inflation. However, as already mentioned in Sec. \ref{Intro}, $\delta s$ becomes ill-defined in cases where the effective mass of isocurvature perturbations would become tachyonic since it would signal a destabilisation effect which is unphysical given that the background evolution remains stable \cite{Cicoli:2018ccr,Cicoli:2019ulk}. As we will see in Sec. \ref{sec:SpuriousInstability}, in this case the proper entropy variable to be used is $\hat{S}_{ij}$.

\begin{figure}[H]
\begin{center}
\includegraphics[width=0.95\textwidth]{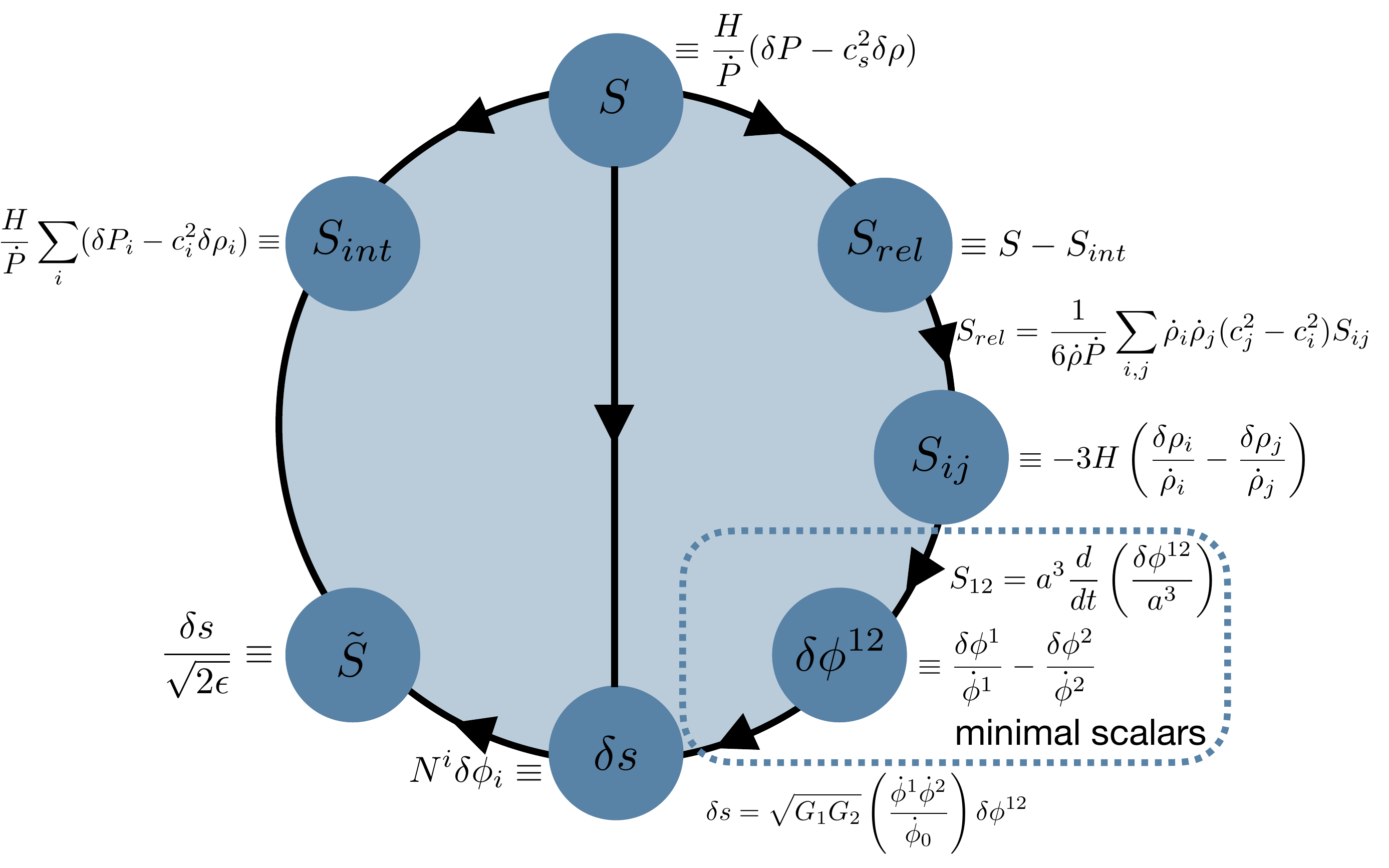}
\caption{Summary of various entropy perturbation variables found in the literature and relations between them. The relations contained within the dashed rectangle hold only for `minimal scalars', i.e. scalar fields with sum-separable potentials and canonical kinetic terms.}
\label{fig:entropy}
\end{center}
\end{figure}

\subsubsection*{Minimal scalars}

For minimally coupled scalars with canonical kinetic terms and sum-separable potentials, which we will call `minimal scalars', one can show that the relative entropy perturbation takes the form:
\be
S_{\rm rel}=\frac{1}{3 \dot\phi_0^2} \frac{V_N \dot\phi^1 \dot\phi^2}{3 H \dot\phi_0+2 V_T }\,S_{12}\,,
\label{eq:Srel}
\ee
where $V_T=T^i V_i$ and $V_N=N^i V_i$ are respectively the projections of the scalar potential along the directions tangent and orthogonal to the background trajectory defined in (\ref{eq:TN}). Using the notation of \cite{Hwang:2000jh}, the relative entropy variable $S_{12}$ can, in turn, be expressed as:
\be
S_{12}=a^3 \frac{d}{d t }\left(\frac{\delta \phi^{12}}{a^3}\right)\,,
\label{eq:S12}
\ee
where $\delta \phi^{12}$ is the generalised entropy introduced in (\ref{dphi12}). In the simple case of a potential of the form $V = \frac12 m_1^2(\phi^1)^2 + \frac12 m_2^2(\phi^2)^2$, in the slow-roll approximation (\ref{eq:S12}) reduces to $S_{12}\simeq -3H \delta\phi^{12}$, which using (\ref{eq:deltas}) with $G_1=G_2=1$ gives:
\be
\tilde{S} \simeq -\frac{\dot{\phi}^1\dot{\phi}^2}{3\dot\phi_0^2}\,S_{12}\,,
\ee
showing that in this particular case $\tilde{S}$ becomes proportional to $S_{12}$. Let us stress that in the general case $\tilde S$ is proportional only to the total entropy perturbation, as in (\ref{Sds}), and not to the relative entropy perturbation. 

The relations between the different definitions of entropy perturbations discussed above are summarised pictorially in Fig. \ref{fig:entropy}. We stress in particular which of them can be used only in case of flat field space without kinetic coupling.

\subsection{Entropy perturbations after inflation}
\label{sec:PerturbationsAfterInflation}

After inflation and the subsequent reheating phase, the Universe is expected to enter a radiation dominated epoch where photons, baryons, cold dark matter and neutrinos make up the primordial plasma. Neglecting velocity isocurvature perturbations, the presence of entropy perturbations at this stage leads to a difference in the number density perturbations, $\delta n_i/n_i$, between the various species. Taking photons $\gamma$ as a reference, we can thus have different non-vanishing relative entropies $S_{n\gamma}$, where $n$ can be cold dark matter, baryons or neutrinos. These are the quantities that are constrained by CMB data.

In fact, the study of cosmological perturbations generated during inflation boils down to the study of CMB anisotropies that can be characterised by their power spectra:
\be
(2\pi)^3 \delta(\mathbf{k} + \mathbf{q}) P_{IJ}(k) = \langle I(\mathbf{k}) J(\mathbf{q})\rangle \,,
\ee
where $I$ and $J$ can denote the curvature $\zeta = \sum_i \zeta_i$ or any post-inflationary isocurvature perturbation $S_{n\gamma}$. Observational constraints can then be formulated in terms of the `primordial isocurvature fraction':\footnote{We are not interested in this paper on the other parameter that is constrained by Planck observations, namely the `correlation fraction':
\be
\cos\Delta_{IJ} = \frac{P_{IJ}}{(P_{II} P_{JJ})^{1/2}} \in (-1, 1) \,,
\ee
which is taken to be scale-independent in Planck analyses \cite{Akrami:2018odb}.}
\be
\beta_{\rm iso}(k) = \frac{P_{S_{n\gamma}}(k)}{P_\zeta (k) + P_{S_{n\gamma}}(k)} \,, 
\ee
where $P_{S_{n\gamma}}(k)\equiv P_{S_{n\gamma}S_{n\gamma}} (k)$ and $P_\zeta(k)\equiv P_{\zeta \zeta}(k)$. In general this quantity is not scale invariant, so that Planck constraints are placed at three different reference scales, with the observational upper bound on $\beta_{\rm iso}$ going from about $1\%$ to $10\%$ depending on the species involves \cite{Akrami:2018odb}. Given our illustrative purposes though, we adopt the assumption of~\cite{Montandon:2020kuk} and take the spectral index for all the spectra to be zero, so that the primordial isocurvature fraction turns out to be scale-independent. Let us also stress that the constraint on $\beta_{\rm iso}$ comes indirectly: once we have the primordial power spectra $P_{IJ}$ at the start of the radiation dominated era (that implies that we have consistently transferred all perturbations from the inflationary to the post-inflationary era) we need to evolve them through the Einstein equations till the release of the CMB, and then translate them into observable quantities using the Boltzmann equations. Only then we are able to place constraints on $\beta_{\rm iso}$ by means of CMB constraints.

After this discussion of observational constraints, it is clear that we should use the relative entropy variable $S_{ij}$ defined in \eqref{eq:Sab} to transfer the entropy mode from the inflationary scalar field system to the primordial plasma that gets formed after reheating and consists of fluids only. In fact, $S_{ij}$ has already been used in several previous works~\cite{Polarski:1994rz, Langlois:1999dw, Bartolo:2001rt}. In general, in order to make contact between the inflationary and the reheating phase, one needs to have a complete model where the couplings of the inflationary scalars to the various species are known, so that $S_{ij}$ can be evolved up to radiation domination.

Notice instead that we would obtain a result which could be wrong by several orders of magnitude if we used the total entropy $S$ given by (\ref{TotS}) \cite{Huston:2011fr, Huston:2013kgl} since $S$ contains both intrinsic and relative contributions. Similar considerations would apply to the entropy perturbation $\tilde S$ orthogonal to the background trajectory given in (\ref{eq:RSTN}) since (\ref{Sds}) shows that $S$ is proportional to $\tilde S$. Moreover, as we will discuss in more detail in Sec. \ref{sec:SpuriousInstability}, in some specific models \cite{Wands:2000dp,Cicoli:2018ccr, Cicoli:2019ulk} the use of \eqref{eq:RSTN} can lead to an unphysical instability of the entropy perturbations due to the spurious divergence of the normal projector $N^i$. In these cases, however, also the relative entropy $S_{ij}$ itself becomes an inappropriate variable since the curvature $\zeta$ turns out to be ill-defined. One has therefore to use the `improved' relative entropy variable $\hat{S}_{ij}$ given in (\ref{hatSij}) which remains finite during inflation and can be safely matched to the post-inflationary epoch, leading to $\hat{S}_{n\gamma}$ for different species denoted by $n$. In order to compare with observational constraints on $\beta_{\rm iso}$, one has then to infer $S_{n\gamma}$ from $\hat{S}_{n\gamma}$ using again (\ref{hatSij}). Notice that there is no apparent divergence in the radiation dominated era since $S_{n\gamma}$ is a well-defined quantity.

\section{Spurious instability from light kinetically coupled fields}
\label{sec:SpuriousInstability}

In this section we shall discuss in depth cases where $S_{ij}$ and $\tilde S$ become ill-defined. In the case of kinetically coupled light scalars, the authors of \cite{Cicoli:2019ulk,Cicoli:2018ccr} found that the isocurvature perturbation $\tilde S$ grows rapidly on super-horizon scales despite the fact that the background trajectory is essentially single field and stable. We will now show that this apparent paradox is caused by the use of wrong entropy variables. 

\subsection{Evolution of isocurvature perturbations}

Before analysing a particular toy-model, let us briefly review how to describe the evolution of isocurvature perturbations in generic multifield models with a curved field space. We will first consider the `field space basis' where scalar perturbations are parametrised as $\delta \phi^i$, and then the `kinematic basis' where the perturbations are decomposed in modes tangent and orthogonal to the background trajectory. 

\subsubsection*{Field space basis}

The gauge invariant scalar perturbations defined as $Q^i=\delta\phi^i + \Psi \dot{\phi}^i/H$ follow \cite{Sasaki:1995aw}:
\be
D_t^2 Q^i+3H D_t Q^i-a^{-2}\,\nabla^2 Q^i+{C^i}_j Q^j=0 \,,
\label{eq:SasakiStewart}
\ee
where $C^i_j$ is given by:
\be
C^i_j=\nabla_j V^i-\dot{\phi}^2 R^i_{k\ell j} T^k T^\ell +2 \epsilon\,\dot\phi_0^{-1}\,H\left(T^i V_j+T_j V^i\right)+2\epsilon(3-\epsilon)H^2 T^i T_j\,.
\ee
In order to analyse the stability of these perturbations it is useful to expand the covariant derivatives, thereby recasting the Mukhanov-Sasaki (MS) equation into the following form:
\be
\ddot{Q}^i+\dot{Q}^j \left(2\Gamma^i_{jk}\dot{\phi}^k+3 H \delta^i_j\right)-a^{-2}\,\nabla^2 Q^i+{ (\mathbb{M}^2)^i}_j Q^j=0\,,
\label{eq:MSQa}
\ee
where the mass-squared matrix looks like:
\be
{(\mathbb{M}^2)^i}_j\equiv C^i_j+\left(\Gamma^i_{jk,\ell}+\Gamma^i_{\ell m}\Gamma^m_{jk}\right)\dot{\phi}^k\dot{\phi}^\ell+\Gamma^i_{jk}\left(\ddot{\phi}^k+3 H\dot{\phi}^k\right).
\ee
Making use of the background equations of motion and recalling that:
\be
{R^i}_{jk\ell}\equiv \Gamma^i_{j\ell,k}-\Gamma^i_{jk,\ell}+\Gamma^i_{km}\Gamma^m_{\ell j}-\Gamma^i_{\ell m}\Gamma^m_{kj}\,,
\ee
the mass-squared matrix can be simplified to:
\be
{(\mathbb{M}^2)^i}_j=\partial_j V^i+\Gamma^i_{k\ell,j}\dot{\phi}^k\dot{\phi}^\ell+2\epsilon\,\dot\phi_0^{-1}\,H\left(T^i V_j+T_j V^i\right)+2\epsilon(3-\epsilon) H^2 T^i T_j\,.
\label{Msqfb}
\ee
Should there be an instability, at least one of the eigenvalues of $\mathbb{M}^2$ would be large and negative in order to overcome the friction term in (\ref{eq:MSQa}) and to drive the growth of the perturbations on super-horizon scales.

\subsubsection*{Kinematic basis}

As already mentioned in Sec. \ref{sec:PerturbationsDuringInflation} regarding the entropy field $s$, in multifield setups it is often useful to work in the kinematic basis rather than in the field space one. One projects the field space perturbations $Q^i$ onto the kinematic basis by using the vielbeins $e^I_i$:
\be
Q^I={e^I}_i \,Q^i\,,
\label{eq:QI}
\ee
allowing to write \eqref{eq:MSQa} in this new basis as \cite{Achucarro:2010da}:
\be
\mathbb{D}_t^2 Q^I +3H \mathbb{D}_t Q^I-a^{-2}\,\nabla^2 Q^I+{C^I}_J Q^J=0\,,
\label{eq:pertkin}
\ee
where ${C^I}_J =e^I_i e^j_J C^i_j$ and the new covariant derivative is defined as $\mathbb{D}_t Q^I\equiv \dot{Q}{^I}+Y^I_J Q^J$ with $Y^I_J\equiv{e^I}_i D_t{e^i}_J $. In order to analyse the stability of this set of coupled oscillators it is useful, as before, to expand the covariant derivatives and to write the MS equation as:
\be
\ddot{Q}^I+(3 H \delta^I_J+2 {Y^I}_J)\dot{Q}^J-a^{-2}\,\nabla^2Q^I+(\mu^2)^I_J Q^J=0\,,
\label{MSeq}
\ee
where we defined the mass-squared matrix of the perturbations in the kinematic basis as:
\be
{\left(\mu^2\right )^I}_J\equiv {C^I}_J+{{\dot{Y}}^I }_{\:\: J}+3H {Y^I}_J+{Y^I}_K{Y^K}_J\,.
\ee
In a simple 2-field case one may choose: 
\be
e^i_1=e_T^i=T^i \qquad\text{and}\qquad e^i_2=e_N^i=N^i\,,
\ee
where $T^i$ and $N^i$ are the tangent and normal vectors to the background trajectory introduced in (\ref{eq:TN}). The background dynamics implies:
\be
D_t T^i=-H \eta_\perp N^i \qquad\text{and}\qquad D_t N^i=H \eta_\perp T^i\,,
\ee
where $\eta_\perp = V_i N^i/(H\dot\phi_0)$ is the turning rate of the trajectory. These relations yield: 
\be
{Y^T}_T={Y^N}_N=0 \qquad \text{and} \qquad {Y^T}_N=-{Y^N}_T= H \eta_\perp\,.
\ee
One can then show that: 
\bea
{\left(\mu^2\right )^T}_T&=&V_{TT}+4\epsilon \dot\phi_0^{-1} H V_T+2 \epsilon(3-\epsilon) H^2-(H \eta_\perp)^2\,, \label{first} \\
{\left(\mu^2\right )^N}_N&=&V_{NN}+\dot\phi_0^2\, R-(H \eta_\perp)^2\,, \label{muNN} \\
{\left(\mu^2\right )^T}_N&=&V_{NT}+ 2\epsilon \dot\phi_0^{-1} H V_N+H^2\eta_\perp (3-\epsilon -\xi_\perp)\,,
\label{Asympt}
\eea
where $R$ is the Ricci scalar of the field space and $\xi_\perp \equiv -\dot\eta_\perp/(H \eta_\perp)$.

\subsection{A 2-field model with an apparent instability}
\label{2fieldModel}

In order to illustrate our claims in the simplest possible terms, in what follows we shall focus on a 2-field system, $(\phi^1,\phi^2)$, where the metric and the scalar potential look like:
\be
G_{ij}= \left(
  \begin{array}{cc}
   1 & 0  \\
0 & f^2(\phi^1) \\
  \end{array} \right)\qquad \text{and} \qquad V = V(\phi^1,\phi^2) \,.
	\label{simplemodel}
\ee
It can be shown that the field space described by (\ref{eq:L}) with (\ref{simplemodel}) exhibits a non-vanishing scalar curvature that takes the form:
\be
R=-2 \,\frac{f_{11}}{f}\,,
\label{eq:R}
\ee
where $f_i\equiv \partial_{\phi^i} f$. The equations of motion become:
\be
\begin{array}{ll}
\Box\phi^i +g^{\mu\nu}\Gamma_{jk}^i\partial_\mu{\phi}^j\partial_\nu\phi^k+G^{ij} V_j=0\ ,
\end{array}
\ee
where $\Box \phi^i=\frac{1}{\sqrt{|g|}}\partial_\mu\left(\sqrt{|g|}g^{\mu\nu }\partial_\nu\phi^i\right)$ and $\Gamma^i_{jk}$ are Christoffel symbols which describe the metric connection of the field space. 

\begin{figure}[!t]
\begin{center}
\includegraphics[width=0.33\textwidth]{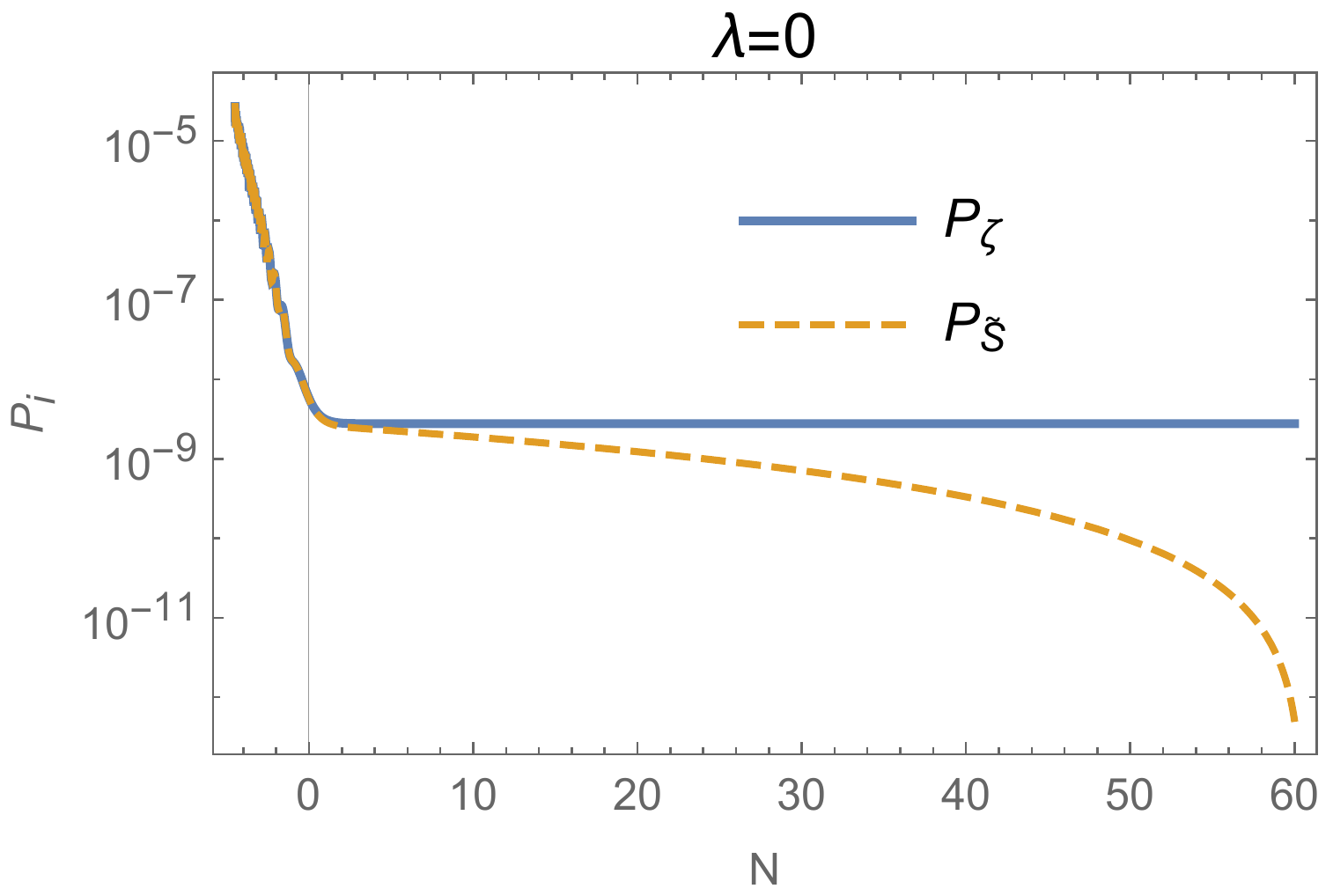}\includegraphics[width=0.325\textwidth]{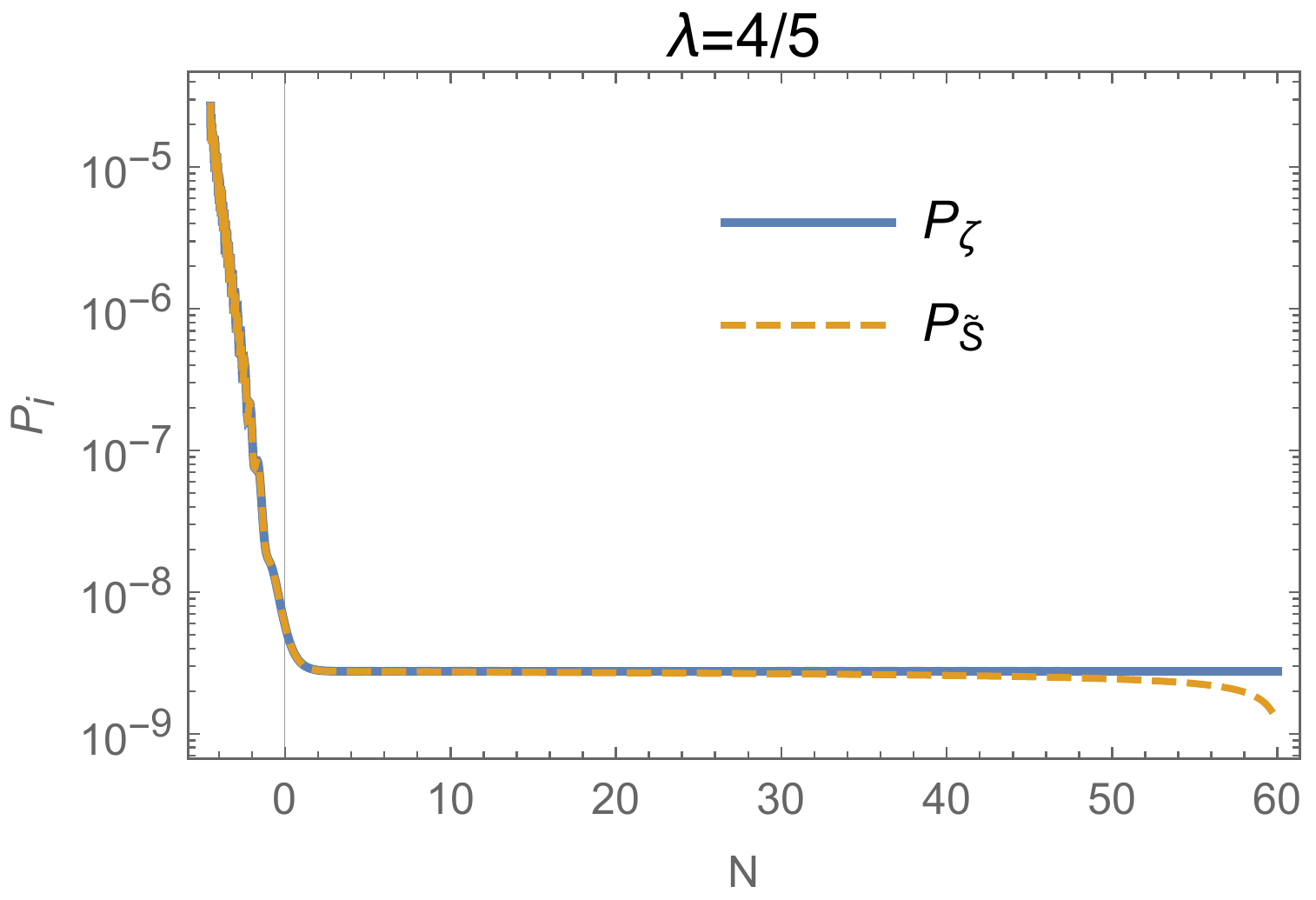}\includegraphics[width=0.335\textwidth]{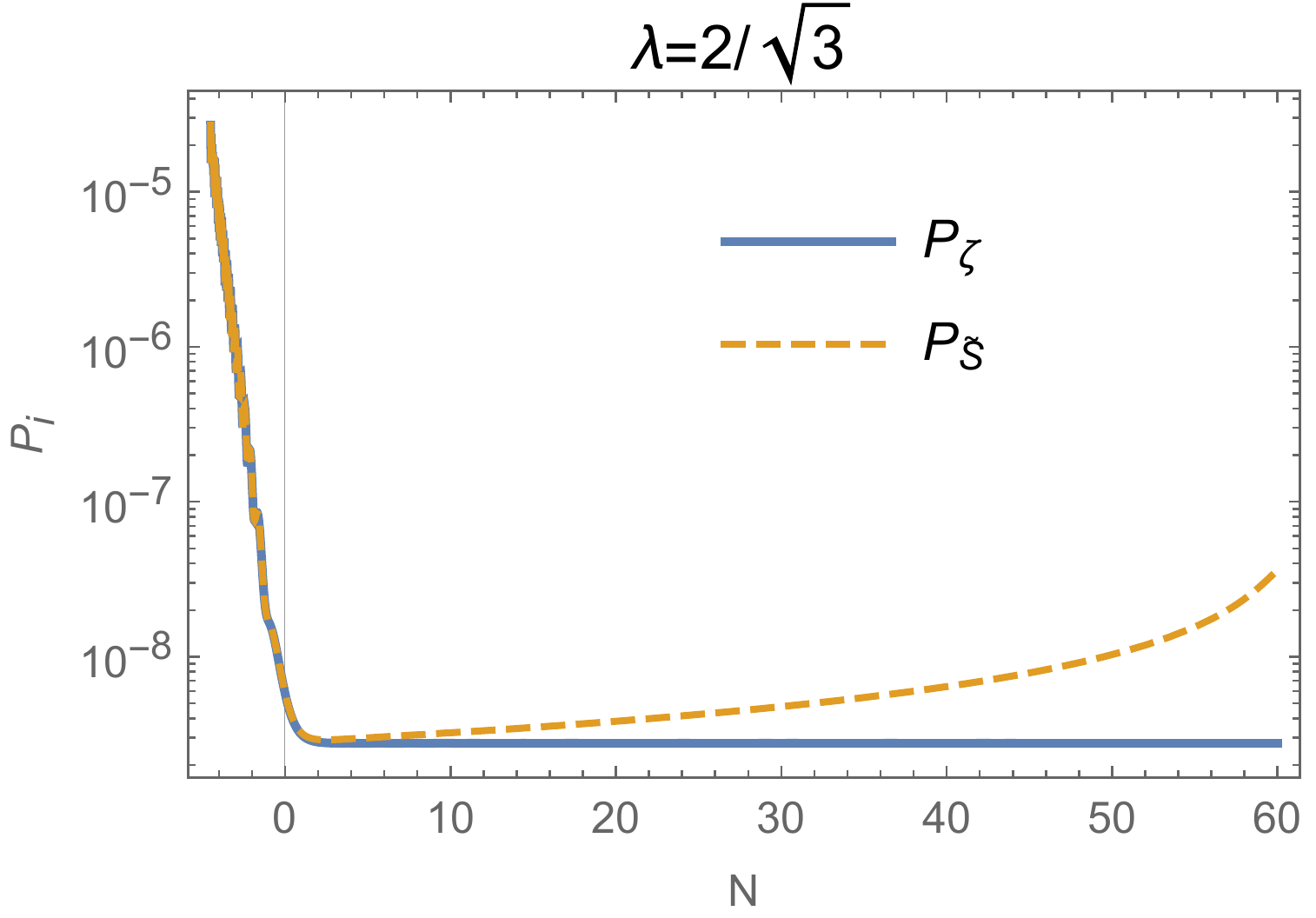}
\caption{Adiabatic and isocurvature power spectra (in terms of $\tilde{S}$) for the system described by (\ref{eq:example}) and (\ref{eq:kc_example}) with $m_2=0$. Left: for $\lambda=0$ the isocurvature perturbations decay on super-horizon scales. Centre: for $\lambda=4/5$ the mass of the isocurvature perturbations is tachyonic but the isocurvature power spectrum still decays because of Hubble friction. Right: for $\lambda=2/\sqrt{3}$ we have tachyonic isocurvature modes that produce an exponential growth of $\tilde{S}$.}
\label{fig:isogrowth}
\end{center}
\end{figure}

These models may give rise to non-geodesic motion in field space with a curved trajectory. Considering homogeneous fields $\phi^i=\phi^i(t)$ in an expanding Universe with $\sqrt{|g|}=a^3$, the equations of motion can be written as:
\be
\dot\pi_1 = a^3 \left(f f_1(\dot\phi^2)^2 -V_1\right), \qquad 
\dot\pi_2 =-a^3 V_2\,,
\label{eq1}
\ee
where $\pi_i$ are the conjugate momenta ($\pi_i=\partial \mathcal{L}/\partial \dot{\phi}^i$) given by:
\be
\pi_1 = a^3 \dot\phi^1\,, \qquad\qquad \pi_2 = a^3 f^2 \dot\phi^2\,.
\label{eq2}
\ee
The background dynamics of the system is determined by (\ref{eq1}), (\ref{eq2}) and the Friedmann equation:
\be
3 H^2= \frac12 G_{ij}\dot{\phi}^i\dot{\phi}^j+V \,.
\ee
In order to understand the effects of the kinetic coupling on the inflationary dynamics we need to analyse both the background trajectory and cosmological perturbations. To provide some quantitative results, let us assume for concreteness that the potential for the inflaton $\phi^1$ is of the Starobinsky form \cite{Starobinsky:1980te} and the spectator field $\phi^2$ is equipped with a quadratic potential:
\be 
V(\phi^1,\phi^2)=\Lambda^4 \left(1-e^{\sqrt{\frac23}\,\phi^1}\right)^2 +\frac{m_2^2}{2}\,(\phi^2)^2\,,
\label{eq:example}
\ee
where we will assume $m_2\ll m_1 = \frac{2}{\sqrt{3}}\,\frac{\Lambda^2}{M_p} < H$ (reinstating appropriate factors of the reduced Planck mass) with $\Lambda\simeq 1.8\times 10^{-10}\,M_p$ in order to match observational bounds on adiabatic perturbations. Moreover we shall consider an exponential kinetic coupling (which is very natural in supergravity and string models) of the form:
\be 
f(\phi^1)=e^{-\lambda \phi^1}\,,
\label{eq:kc_example}
\ee
so that the field space has a constant scalar curvature $R=-2\lambda^2$. The super-horizon behaviour of the isocurvature modes depends on the geometry of the field space. In particular, considering the case of a massless spectator field, i.e. $m_2=0$, the mass associated to the isocurvature perturbations $\delta s$ given by (\ref{muNN}) becomes \cite{Cicoli:2019ulk}: 
\be 
{\left(\mu^2\right )^N}_N=-\lambda V_1 -2 \lambda^2 \epsilon H^2 \,,
\ee
where the first term comes from the field space metric connection (and it is of order $\sqrt{\epsilon}$), while the second from the field space curvature (and it is of order $\epsilon$). Notice that in this case $\eta_\perp=0$. Therefore, if the field space shows an $\sim\mc{O}(1)$ scalar curvature (as typical of supergravity and string models), the sign of $(\mu^2)_N^N$ during inflation is determined by the relative sign between $\lambda$ and the first derivative of the inflationary potential. Given that in Starobinsky inflation $V_1>0$, ${\left(\mu^2\right)^N}_N<0$ for $\lambda>0$.\footnote{Notice that the form of the inflationary potential is crucial to induce a tachyonic entropy perturbation $\delta s$ since a quintessence-like potential of the form $V = V_0\,e^{-k\phi^1}$ would not induce any negative eigenvalue of the mass-squared matrix \cite{Cicoli:2019ulk, Cicoli:2020cfj, Cicoli:2020noz}.} It is important to stress that a negative $(\mu^2)_N^N$ is not sufficient to trigger the growth of the isocurvature perturbations by itself, since (\ref{eq:MSQa}) and (\ref{MSeq}) also feature a Hubble friction term. We show some examples related to different values of $\lambda$ in Fig. \ref{fig:isogrowth}. These results suggest that, despite the fact that for every value of $\lambda$ the system trajectory becomes effectively single field (as signalled by the fact that $\eta_\perp=0$), there might be a growth of isocurvature perturbations triggered by slow-roll suppressed contributions. This somewhat paradoxical state of affairs requires further investigation. Given our interest in understanding the origin of the apparent isocurvature growth, from now on we will focus on the case with $\lambda=2/\sqrt{3}$.

In Sec. \ref{InstabilityOrigin} we reveal the nature of this spurious instability first by uncovering the origin of the growth of the isocurvature perturbations in the kinematic basis and then by showing that the relative entropy perturbation between the two scalars is well-behaved and vanishingly small during inflation.

\subsection{Origin of the spurious instability}
\label{InstabilityOrigin}

In this section we will analyse the eigenvalues of the mass-squared matrices of the perturbations in the field space and in the kinematic basis, showing that the apparent instability of the isocurvature perturbation $\tilde{S}$ is a feature exclusively of the kinematic basis triggered by $(\mu^2)_N^N<0$. In fact, this instability disappears in the field space basis where the mass-squared matrix has no negative eigenvalue, showing that $\tilde{S}$ is not a physical quantity. We will see that a well-behaved, i.e. both gauge invariant and finite, quantity is the relative entropy perturbation $S_{\rm rel} = \hat{S}_{12}$ defined in (\ref{hatSij}).

\begin{itemize}
\item \textbf{Field space basis}

Restricting our attention to the model described by (\ref{eq:example}) and (\ref{eq:kc_example}) and assuming for simplicity $m_2=0$, one can show that, on-shell, the only non-vanishing entry of the mass-squared matrix (\ref{Msqfb}) is:
\be
{\left(\mathbb{M}^2\right)^1}_1= V_{11}+4\epsilon \dot\phi_0^{-1} H V_1+2\epsilon(3-\epsilon) H^2\,.
\label{eq:mQ22Field}
\ee
As shown in App. \ref{sec:singlefield}, (\ref{eq:mQ22Field}) can be written in terms of the Hubble slow-roll parameters $\epsilon=-\dot{H}/H^2$, $\eta = \dot\epsilon/(\epsilon H)$ and $\kappa = \dot\eta/(\eta H)$ as:
\be
\frac{{\left(\mathbb{M}^2\right)^1}_1}{H^2}=-\frac{3 \eta}{2}+\frac{\epsilon \eta}{2}-\frac{\eta^2}{4}-\frac{\kappa \eta}{2}\,,
\ee
which matches exactly the mass of the $Q$-perturbation obtained in the single field formalism, so excluding the presence of any tachyonic instability in this basis. 

\item \textbf{Kinematic basis}

In the model under study, the relations (\ref{first})-(\ref{Asympt}) reduce asymptotically (i.e. when the system relaxes to the background attractor solution) to:
\bea
{\left(\mu^2\right )^T}_T&=& V_{11}+ 4\epsilon \dot\phi_0^{-1} H\,V_1+2 \epsilon(3-\epsilon) H^2\,, \\
\label{eq:TachyonicMass}
{\left(\mu^2\right )^N}_N&=& \frac{f_1}{f}\,V_1 +\epsilon H^2 R = -\lambda V_1 -2 \lambda^2 \epsilon H^2\,, \\
{\left(\mu^2\right )^T}_N&=&0\,.
\eea
Notice that ${\left(\mu^2\right )^T}_T=(\mathbb{M}^2)^1_1$, an indication that adiabatic perturbations behave the same way in both bases and in a manner compatible with a single field analysis of the model. However, as shown in Sec. \ref{2fieldModel}, we can have a tachyonic orthogonal direction if the metric connection or the field space curvature $R$ give rise to sufficiently negative contributions, signaling the presence of a potential instability. 
\end{itemize}

Given that the difference between the kinematic and the field space basis is the mere multiplication by the vielbein, as in (\ref{eq:QI}), it is natural to conclude that the origin of the apparent geometrical instability is the time evolution of the vielbein, a fact we will now investigate in detail.

\subsubsection*{A growing projector}

In order to pin down the origin of the apparent geometrical instability in the kinematic basis, we consider the time evolution of the vector orthogonal to the background trajectory: 
\be
D_t N^i=H \eta_\perp T^i\,.
\ee
In the cases under consideration the background trajectory is straight, implying that $\eta_\perp=0$, leading to:
\be
\dot N^i =-\Gamma^i_{jk}\dot{\phi}^j N^k\,.
\ee
Furthermore in the attractor solution $\dot\phi^2=0$, implying:
\be
N^1=0\qquad\text{and}\qquad\dot{N}^2= - \Gamma^2_{21} \dot\phi^1 N^2\,.
\ee
The equation for $N^2$ can be integrated to find $N^2=1/f $ or equivalently $N_2=f=e^{-\lambda \phi^1}$, showing that $N_2$ grows during inflation as the inflaton $\phi^1$ rolls down towards the origin of the potential from positive values. Notice that the normalisation of the orthogonal vector, $N^i N_i=1$,  is preserved at all times.\footnote{This time evolution could also have been deduced from the normalisation of the orthogonal vector $N^i$.} Expanding the definition (\ref{deltas}) of the orthogonal perturbation $\delta s$ one finds that it involves only the $N_2$ component:
\be
\delta s = N_1 \delta\phi^1+N_2 \delta\phi^2= N_2 \delta\phi^2\,.
\ee
As shown above, the perturbation $\delta\phi^2=Q^2$ (in the spatially flat gauge) has a vanishing mass, and so it becomes constant on super-horizon scales. On the other hand, as we have just seen, $N^2$ grows during inflation, signalling a potentially dangerous growth of $\delta s$. This is the very origin of the apparent geometric instability observed in \cite{Cicoli:2018ccr,Cicoli:2019ulk}.

However, as mentioned in Sec. \ref{sec:PerturbationsAfterInflation}, observational bounds on isocurvature perturbations are given in terms of the post-inflationary relative entropy perturbation $S_{n\gamma}$ which, after a proper understanding of the reheating epoch, should arise from the inflationary non-adiabatic perturbation $S_{12}$ defined in (\ref{eq:Sab}). Thus the observed growth in $\delta s$ does not rule these models out since the quantity to be considered to match observations is $S_{12}$. However, in this particular case also $S_{12}$ is ill-defined as the action of the spectator field $\phi^2$ is (in the massless limit): 
\be
\frac{\mc{L}}{\sqrt{|g|}}\supset\frac12\,f^2(\phi^1)\, (\partial \phi^2)^2\,,
\ee
which gives $\rho_2=\frac12 \,f^2(\phi^1) (\dot\phi^2)^2$ that rapidly goes to zero during inflation. We therefore conclude that $S_{12}$ is singular. Notice that the same problem would arise for canonically normalised massless scalars (which however do not suffer from a growing $\delta s$).

\subsection{Vanishing relative entropy perturbation}

We now further clarify the spurious nature of the apparent geometric instability which characterises this class of models by showing that the relative entropy $S_{\rm rel}=\hat{S}_{12}$ given by (\ref{eq:Sreldef}) remains finite and vanishingly small during inflation. We argue that $S_{\rm rel}$ provides a consistent definition of isocurvature perturbations that are independent of the (unstable) vielbeins, in order to compute the scalar 2-point functions and to confront the model predictions with observational constraints. 

Let us start the computation of $S_{\rm rel}$ by noticing that the energy-momentum tensor during inflation can be decomposed as:
\be
T^{\mu\nu}=T^{\mu\nu}_{(1)}+T^{\mu\nu}_{(2)}\,,
\ee
where the subscripts $(1)$ and $(2)$ refer to $\phi^1$ and $\phi^2$ respectively. Due to the kinetic coupling, the individual $T^{\mu\nu}_{(i)}$ are not conserved since there is energy transfer between the fluids \cite{Malik:2004tf}:
\be
\nabla_\nu T^{\mu\nu}_{(1)}= \mc{T}^\mu \qquad\text{and}\qquad \nabla_\nu T^{\mu\nu}_{(2)}=-\mc{T}^\mu\,,
\ee
where $\mc{T}^\mu$ is the energy transfer function. Despite the fact that there is some freedom in the definition of the two fluids, it is natural to write the energy and pressure of the two fields as (see also App.~\ref{app:gaugeinvariance}):
\bea
\label{eq:energies}
\rho_1&=&\frac12 (\dot\phi^1)^2+V^{(1)}(\phi^1)\,,\qquad \qquad \rho_2=\frac12 f(\phi^1)^2 (\dot\phi^2)^2+ V^{(2)}(\phi^2)\,,  \\
P_1&=&\frac12 (\dot\phi^1)^2-V^{(1)}(\phi^1)\,,\qquad \qquad P_2=\frac12 f(\phi^1)^2 (\dot\phi^2)^2-V^{(2)}(\phi^2)\,,
\label{eq:pressure}
\eea
where we decomposed the sum-separable potential in (\ref{eq:example}) as $V(\phi^1,\phi^2)=V^{(1)}(\phi^1)+V^{(2)}(\phi^2)$. The energy transfer function takes the form:
\be
\mc{T}^\mu=\begin{pmatrix}
 \dot{\rho}_1+3H(\rho_1+P_1)\\
 \vec{0}\\
\end{pmatrix}=\begin{pmatrix}
f f_1\, \dot\phi^1\, (\dot\phi^2)^2\\
 \vec{0}\\
\end{pmatrix}\,.
\ee
The sound speeds of the two fluids defined in (\ref{eq:energies}) and (\ref{eq:pressure}) are:
\be
c_1^2=1+\frac{2 V_1}{3H\dot\phi^1-f f_1\, (\dot\phi^2)^2}\,,\qquad c_2^2=1+\frac{2 V_2 }{\dot\phi^2 (3Hf^2+ f f_1 \dot\phi^1)}\,,
\ee
while the overall sound speed is given by:
\be
c_s^2=1+\frac{2}{3H}\frac{V_i \dot{\phi}^i}{\dot{\phi}_j\dot{\phi}^j}\,.
\ee
In order to compute the relative entropy perturbation $S_{\rm rel}$, we need to use perturbation theory at linear order. Given that $S_{\rm rel}$ is gauge invariant, we can compute it in any gauge. For simplicity we shall use the spatially flat gauge where the perturbed Einstein equations take the form presented in App.~\ref{AppA}. Energy and pressure perturbations read:
\bea
\delta\rho_1 &=&-\Phi (\dot\phi^1)^2 + \dot\phi^1 \delta\dot\phi^1  +V_1 \delta\phi^1\,, \\
\delta P_1&=&-\Phi (\dot\phi^1)^2 +\dot\phi^1 \delta\dot\phi^1 - V_1 \delta\phi^1\,, \\
\delta\rho_2&=& \delta P_2 = -\Phi f^2 (\dot\phi^2)^2 +f^2 \dot\phi^2 \delta\dot\phi^2 + (\dot\phi^2)^2 f f_1 \delta \phi^1 + V_2\delta\phi^2 \,,
\eea
where Einstein equations imply that the lapse function (defined in App. \ref{AppA}) is given by:
\be
\Phi =\frac{1}{2H} \left(\dot\phi^1 \delta \phi^1 + f^2 \dot\phi^2 \delta\phi^2\right)\,.
\ee
In order to get some analytic results, let us focus again on the massless case with $m_2=0$ (we will however show below also numerical results for the $m_2\neq 0$ case). In the 2-field case, it is easy to see that the relative entropy perturbation (\ref{eq:Sreldef}) can also be written as:
\be
S_{\rm rel}=\frac{H}{\dot P} \left[\left( c_1^2-c_s^2\right)\delta\rho_1+\left( c_2^2-c_s^2\right)\delta\rho_2\right]\,.
\label{Srelnew}
\ee
Before computing this expression, it is important to realise that the equation of motion for $\phi^2$ can be integrated exactly leading to:
\be
\left(f \grave{\phi}^2\right)(N)=\left(f \grave{\phi}^2\right)(0)\;e^{-\int_0^N(3-\epsilon)dN^{'}+\lambda \Delta \phi^1}\,,
\label{eq:standardfphi2}
\ee
where we used the number of e-foldings $N=\int H dt$ as time variable, $\grave{\phi}^2\equiv d \phi^2/dN$ and $\Delta \phi^1=\phi^1(N)-\phi^1(0)$. Given that in Starobinsky inflation $\Delta\phi^1\simeq\sqrt{\frac32}\ln\left(1-\frac{4N}{3}e^{-\sqrt{\frac23}\phi^1(0)}\right)$, it is easy to see that during inflation the velocity of the canonical spectator field becomes rapidly negligible since:
\be
(f\grave{\phi}^2)\simeq (f\grave{\phi}^2)(0)\,e^{-3N} \to 0\,. 
\ee
Using this crucial result, we can now evaluate the different contributions to $S_{\rm rel}$, The terms involving the speeds of sound look like:
\bea
c_1^2-c_s^2&=& (1-c_s^2) (f \grave{\phi}^2)^2\left(\frac{1+\frac{f_1 \grave{\phi}^1}{3f} }{\frac{f_1 \grave{\phi}^1}{3f} (f \grave{\phi}^2)^2-(\grave{\phi}^1)^2}\right)
\sim\mc{O}( (f \grave{\phi}^2)^2)\to 0\,,\\
c_2^2-c_s^2&=&(1-c_s^2)\,,
\eea
while the energy density perturbations can be expressed as:
\bea
\delta\rho_1 &=& H^2 \left[\left(\frac{V_1}{H^2}-\frac{(\grave{\phi}^1)^3}{2}\right)\delta\phi^1+\grave{\phi}^1\delta\grave{\phi}^1\right]- \underbrace{\frac12 H^2 f (f \grave{\phi}^2)\,(\grave{\phi}^1)^2\delta\phi^2}_{\mc{O}( f \grave{\phi}^2)\to 0} \\ 
\delta\rho_2 &=& H^2 (f\grave{\phi}^2) \left[\frac12 (f \grave{\phi}^2) \left(\frac{2 f_1}{f}-\grave{\phi}^1\right)\delta\phi^1+f\delta\grave{\phi}^2-\frac12 f (f\grave{\phi}^2)^2\delta\phi^2\right] \sim\mc{O}( f \grave{\phi}^2)\to 0\,. \nonumber
\eea
Finally the ratio $H/\dot P$ is finite and well-behaved since:
\be
\frac{H}{\dot P} = -\frac{1}{6\epsilon H^2 + 2 V_T \sqrt{2\epsilon} }\,.
\ee
Thus these relations, when inserted in (\ref{Srelnew}), imply that after a few e-foldings of inflation the relative entropy perturbation becomes immediately negligible since $S_{\rm rel}\sim \mc{O}(f\grave{\phi}^2)\rightarrow 0$. 

\begin{figure}[!t]
\begin{center}
\includegraphics[width=0.33\textwidth]{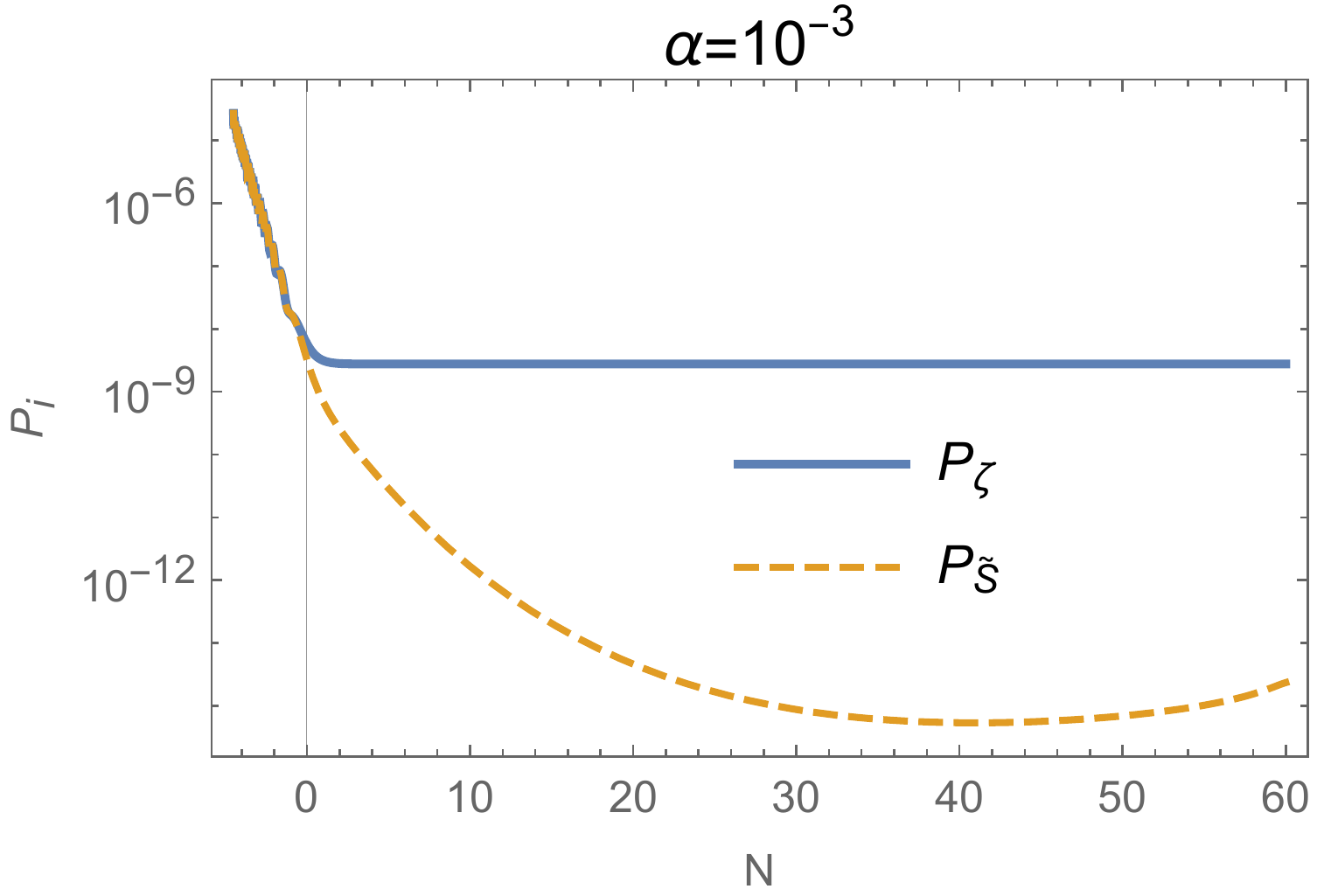}\includegraphics[width=0.33\textwidth]{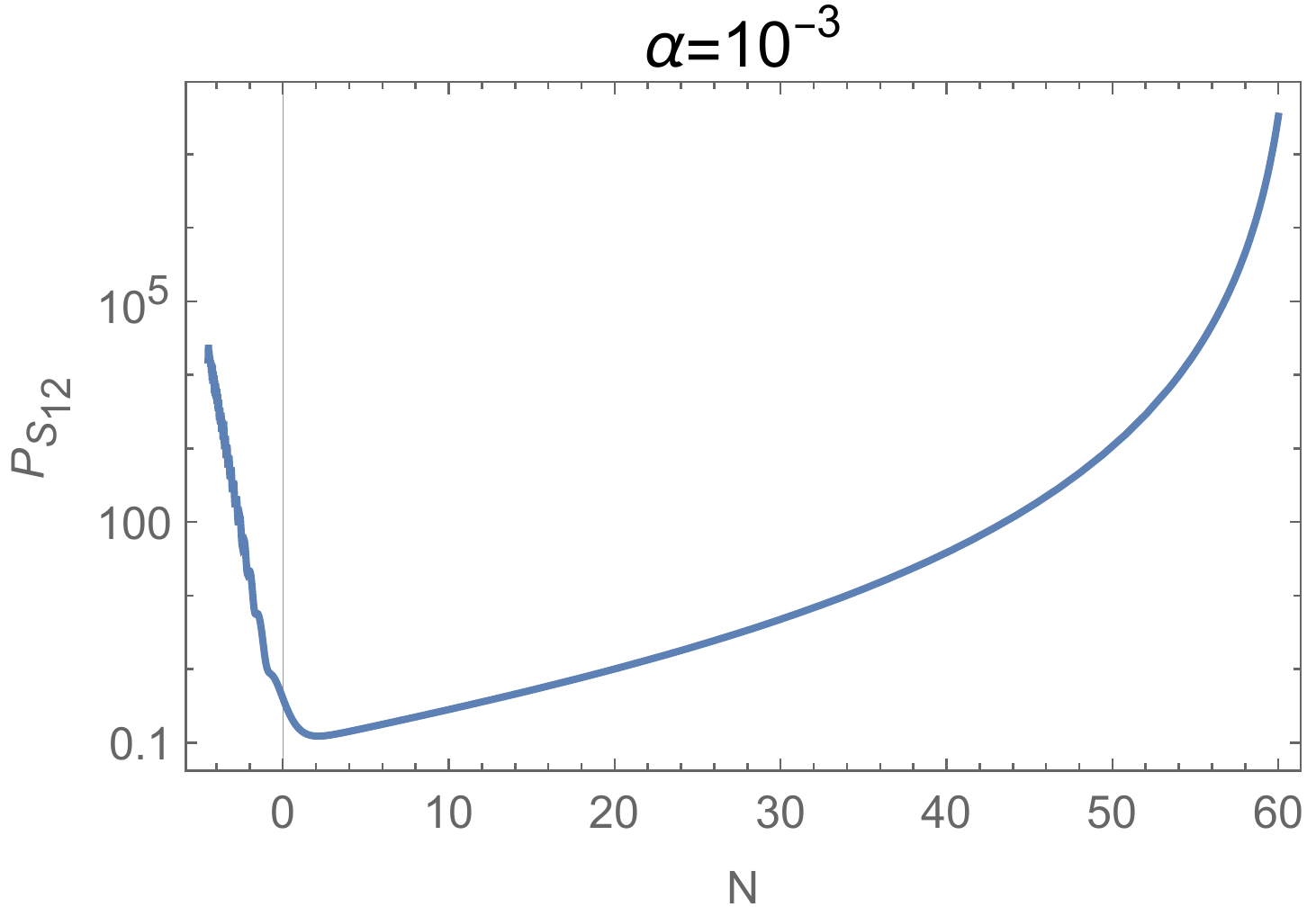}\includegraphics[width=0.33\textwidth]{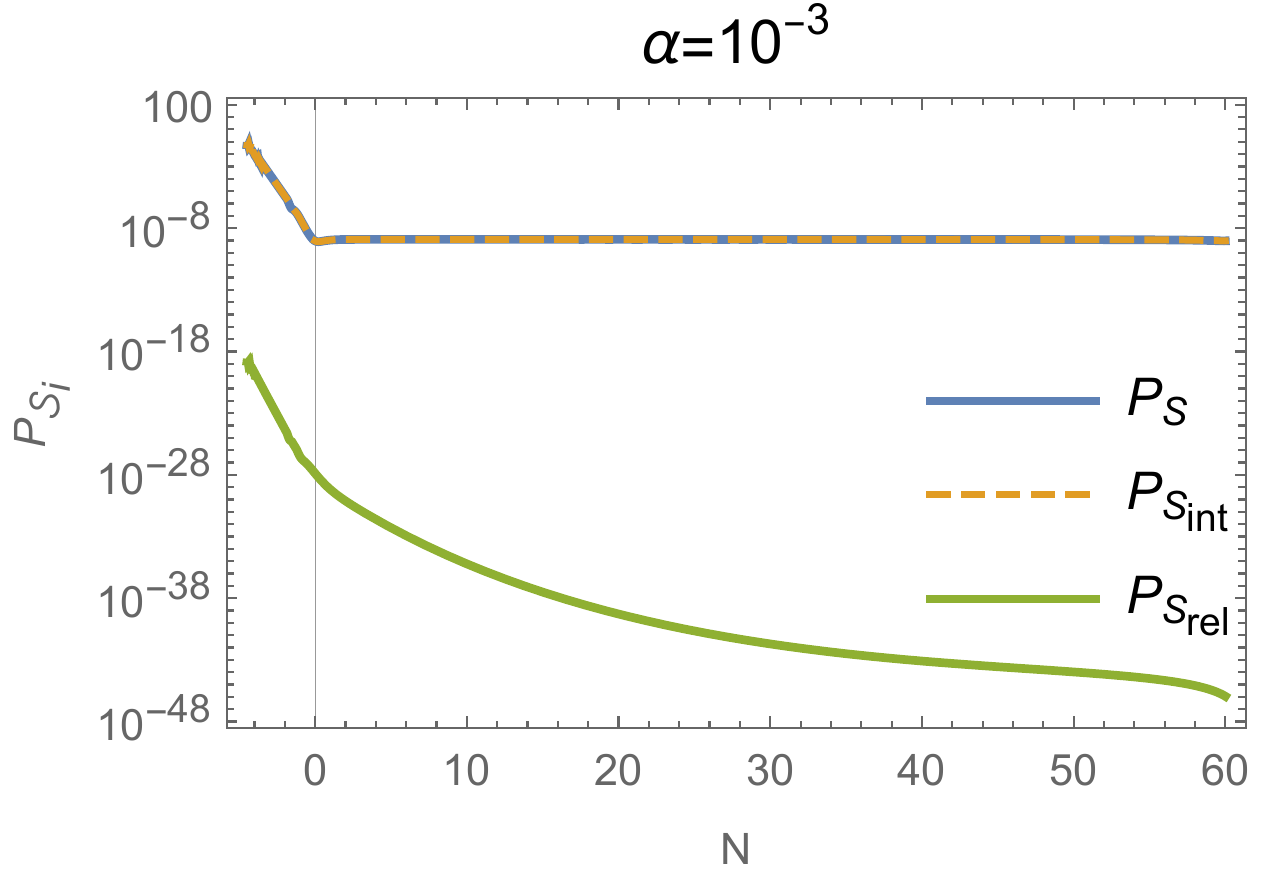}\\
\includegraphics[width=0.33\textwidth]{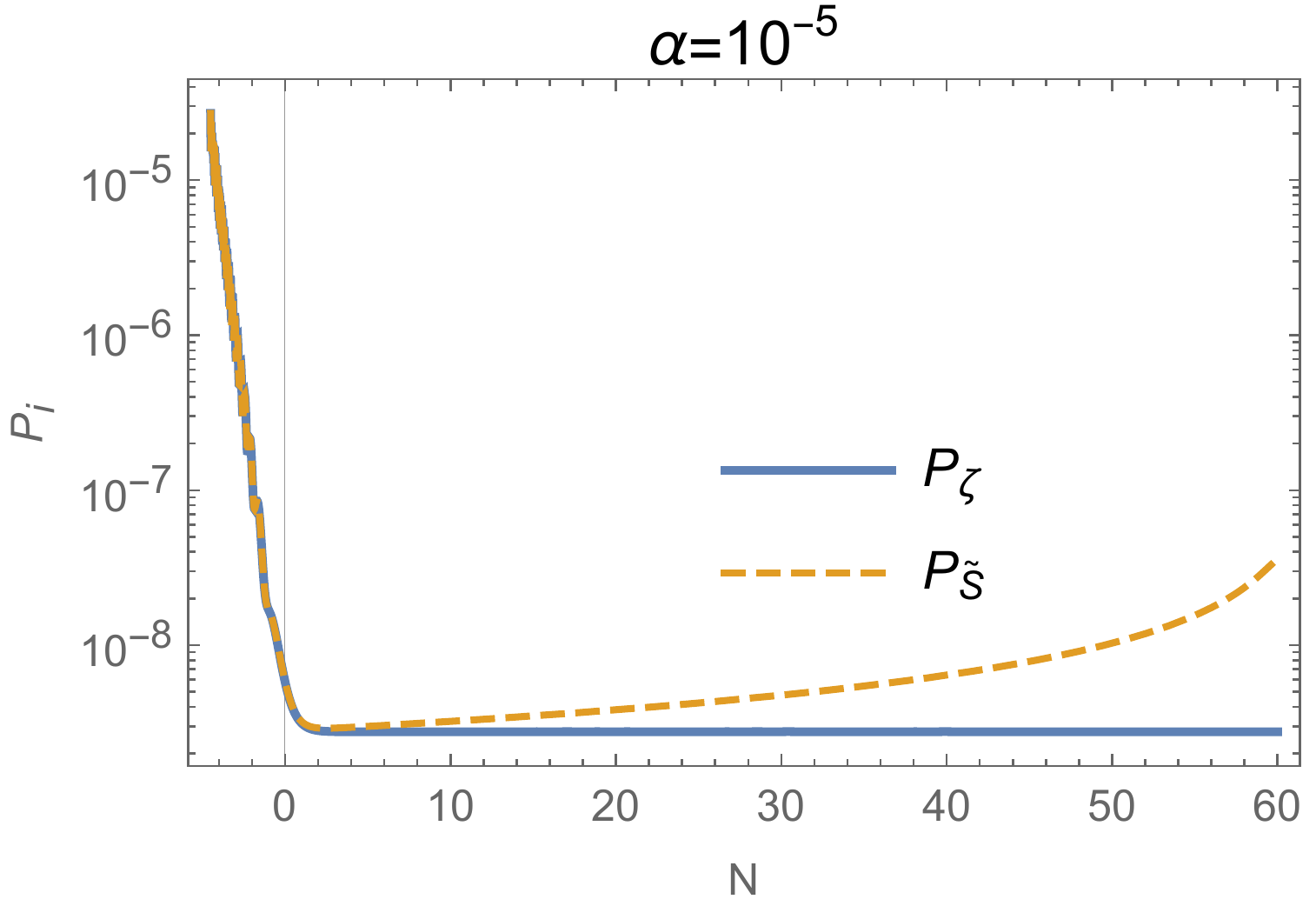}\includegraphics[width=0.33\textwidth]{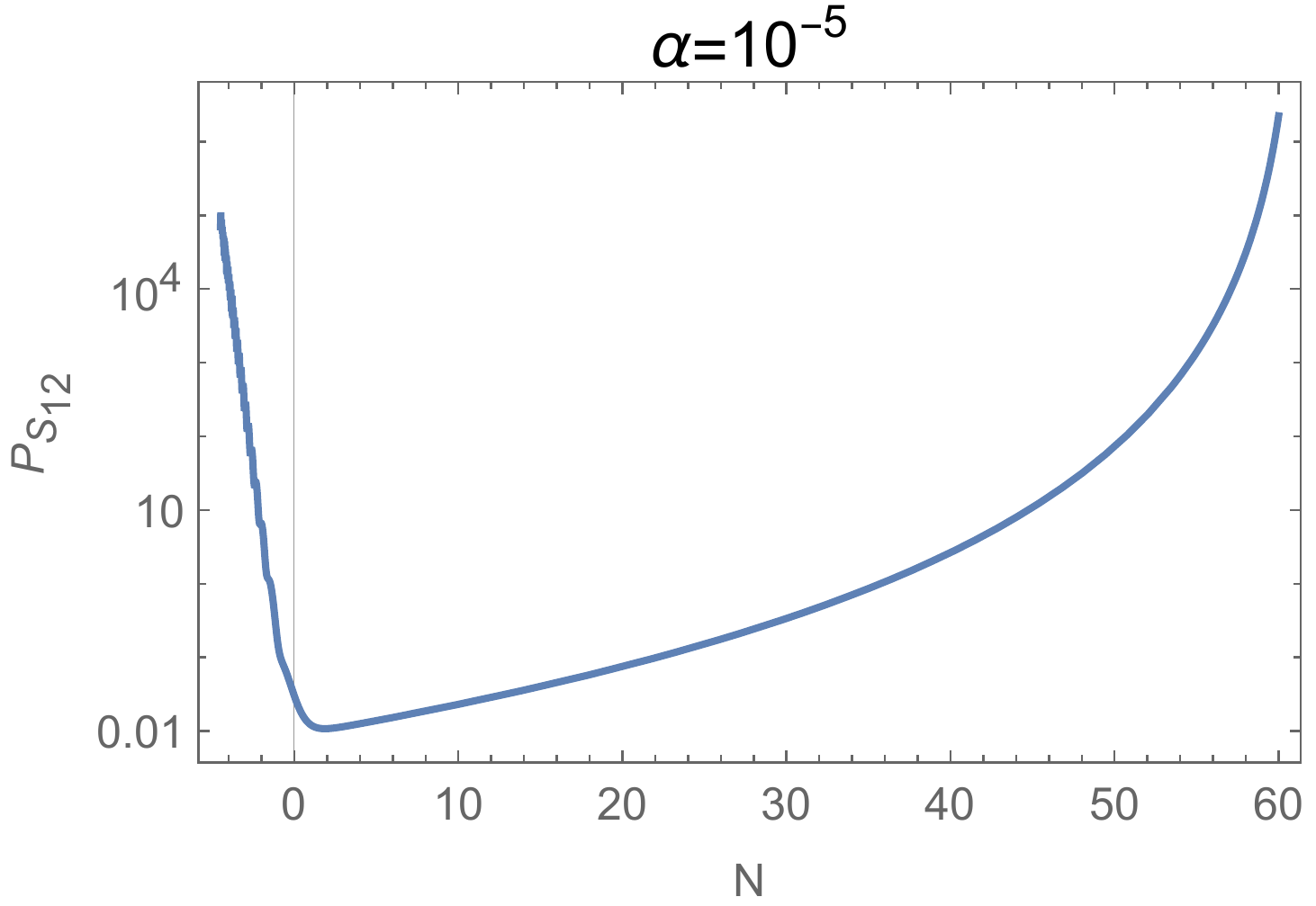}\includegraphics[width=0.33\textwidth]{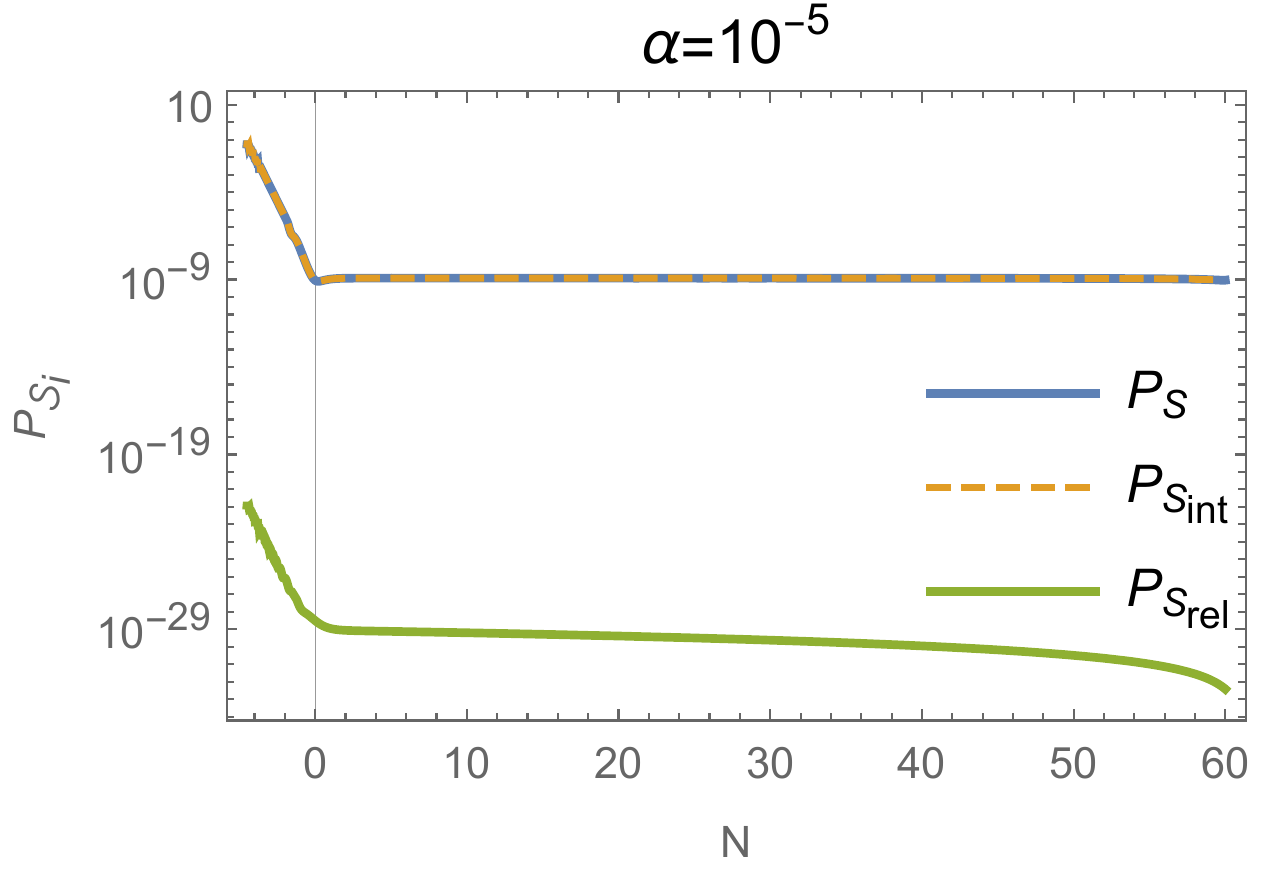}\\
\includegraphics[width=0.33\textwidth]{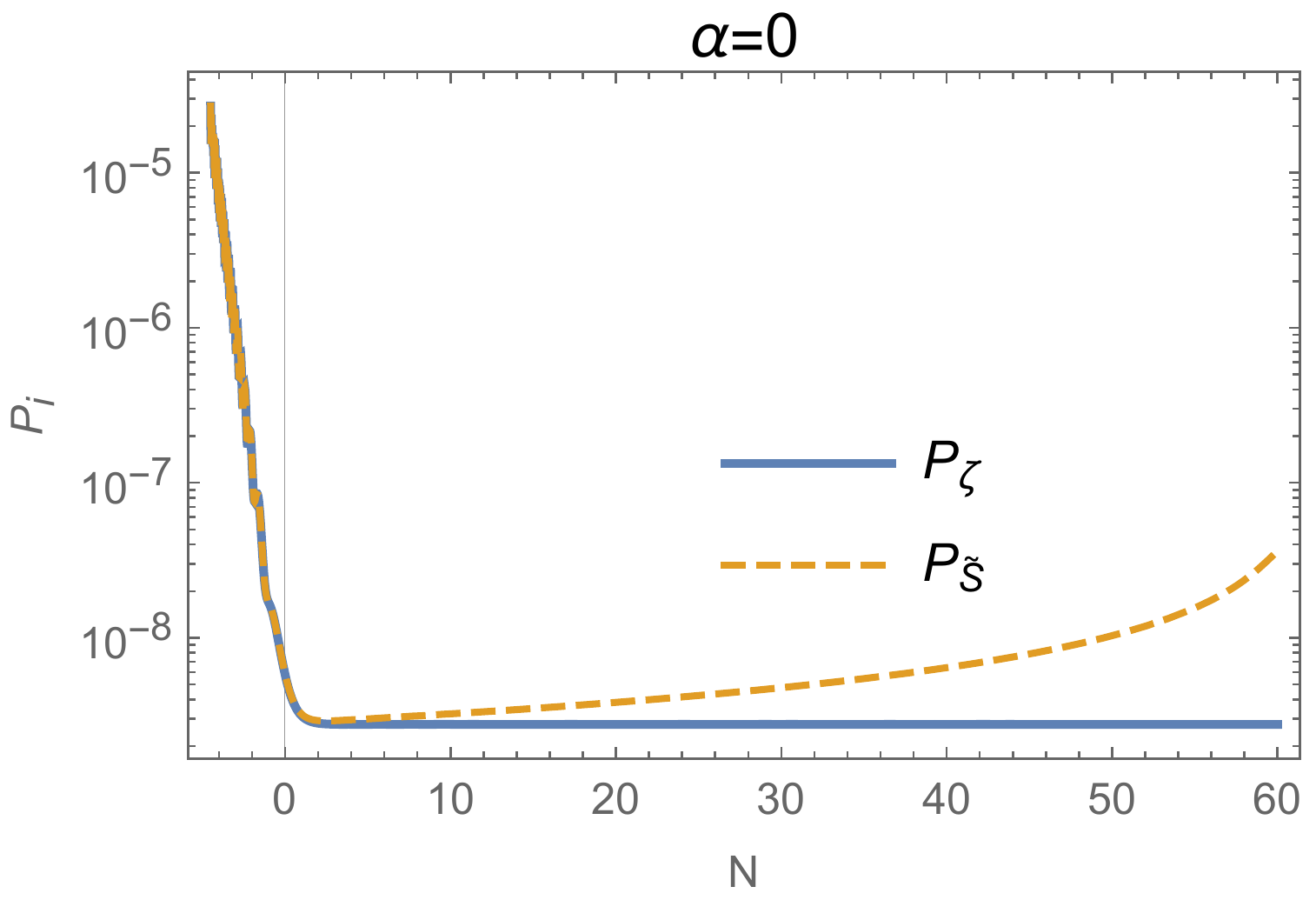}\includegraphics[width=0.33\textwidth]{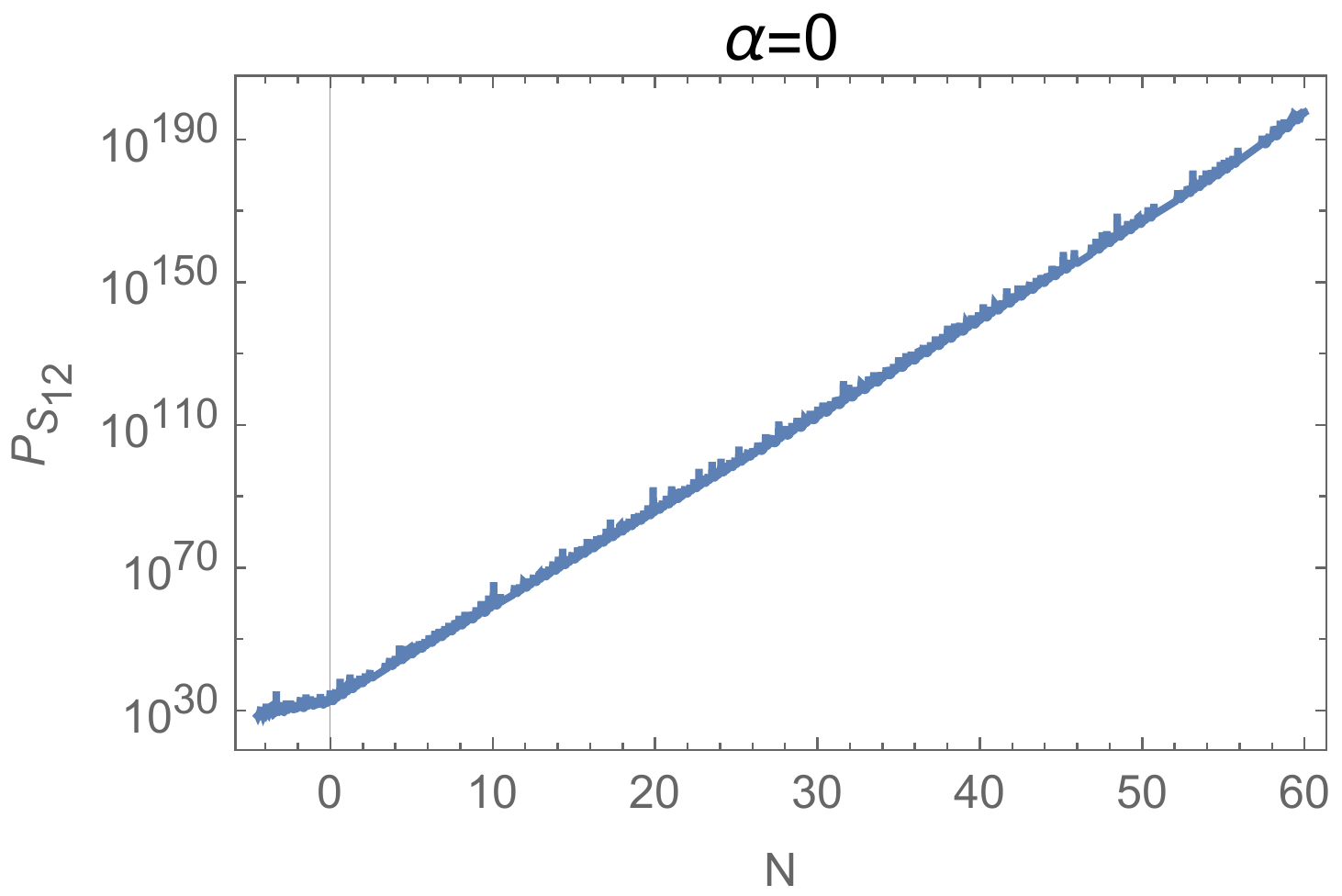}\includegraphics[width=0.33\textwidth]{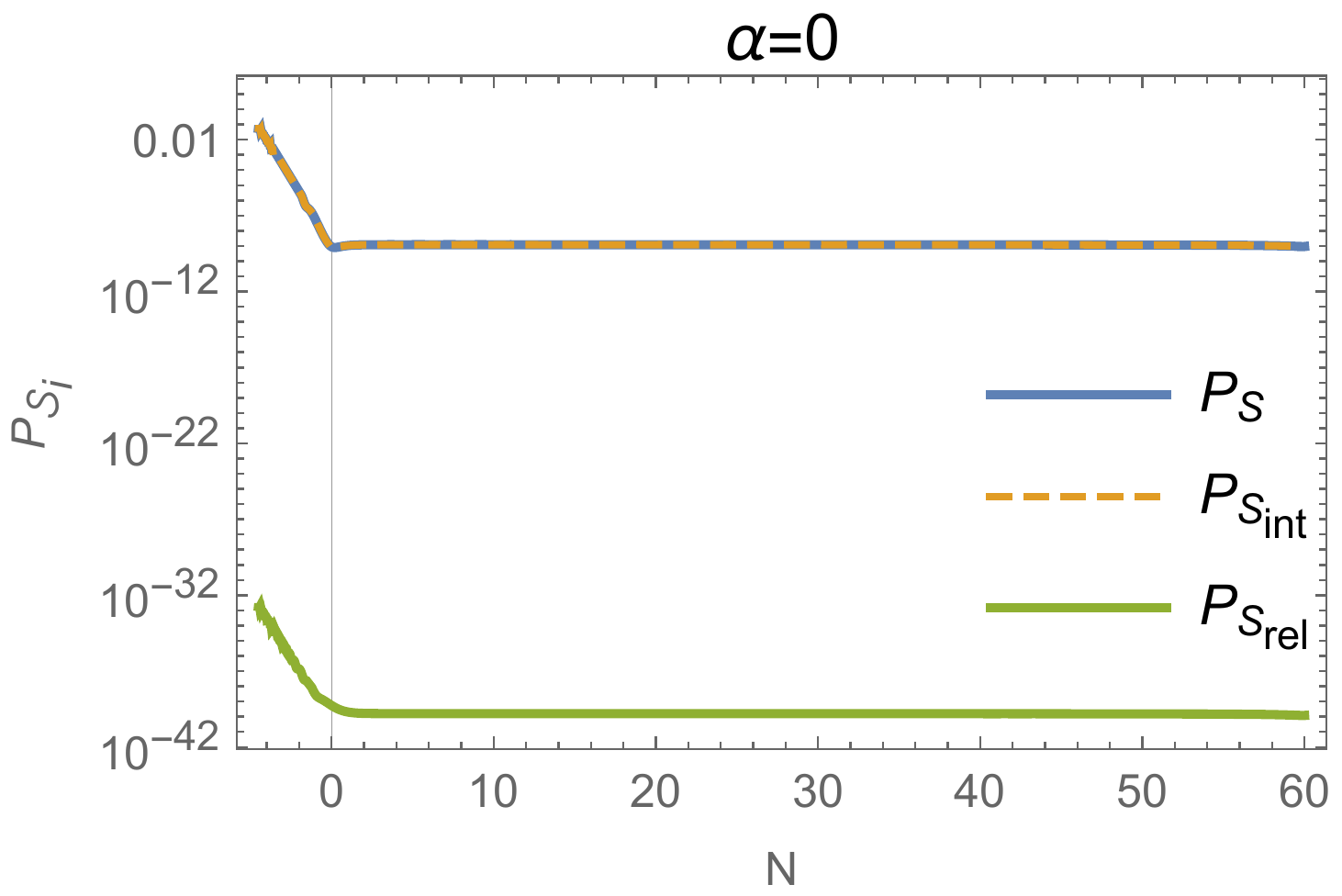}\\
\caption{Time evolution of all entropy perturbation variables for modes exiting the horizon $60$ e-foldings before the end of inflation in the system described by (\ref{eq:example}) and (\ref{eq:kc_example}) for different values of $\alpha=m_2/m_1$. The initial conditions are $\phi^1=5.7$, $\phi^2=0.5$, $\grave{\phi}^1=0$ and $\grave{\phi}^2=0.1$. Left: adiabatic and isocurvature power spectra using the entropy variable $\tilde{S}$ which decays only for $\alpha=10^{-3}$. Centre: spurious growth of isocurvature perturbations using the relative entropy perturbation $S_{12}$. Right: isocurvature power spectrum in terms of the total ($S$), the intrinsic ($S_{\rm int}$) and the relative ($S_{\rm rel}$) entropy perturbation which is decreasing and subdominant on super-horizon scales.}
\label{fig:Scomponents}
\end{center}
\end{figure}

We confirmed this analytic result by performing a numerical analysis of the evolution of all entropy variables $S$, $S_{\rm int}$, $S_{\rm rel}$, $S_{12}$ and $\tilde{S}$ for 3 different values of the parameter $\alpha=m_2/m_1=\{10^{-5},10^{-3},0\}$ which fixes the mass hierarchy between $\phi^2$ and $\phi^1$. The outcome is presented in Fig. \ref{fig:Scomponents} for modes exiting the horizon $N=60$ e-foldings before the end of inflation. Notice that the isocurvature power spectrum computed using $\tilde{S}$ grows not just for $\alpha=0$ but also for $\alpha=10^{-5}$, while it decays for $\alpha=10^{-3}$ even it features a slight climb back up between $50$ and $60$ e-foldings. This implies that the perturbation $\delta s$ orthogonal to the background trajectory is a bad variable not just in the $m_2=0$ case but also more in general for spectator fields lighter than the inflaton. This is true, above all, also for the relative entropy perturbation $S_{12}$ which clearly diverges for all values of $\alpha$ since the energy density $\rho_2$ becomes in practice constant during inflation. On the other hand, the `improved' relative entropy perturbation $\hat{S}_{12}=S_{\rm rel}$ gives rise to a decaying non-adiabatic power spectrum. 

We therefore conclude that the apparent geometrical instability of models with light kinetically coupled spectator fields is unphysical and this class of models, upon exit from inflation, gives rise to a radiation phase where perturbations of the gravitational potential are set by the inflationary curvature perturbations and where there are no isocurvature perturbations, just like in single field inflationary models.

\section{Conclusions}
\label{sec:Conclusions}

Effective field theories characterised by multiple scalar fields and a Riemannian field manifold are a ubiquitous feature of gravitational theories with non-minimal couplings, supergravity models and string theory compactifications. One of their primary applications is early Universe cosmology where the dynamics of the scalar field system drives cosmic inflation. In a multifield framework the study of cosmological perturbations deserves however careful attention. In particular, non-adiabatic fluctuations can potentially rule these models out by giving rise to isocurvature perturbations whose amplitude exceeds observational constraints. Various entropy variables have been proposed so far in the literature, with different pros and cons.

In this paper we clarified which of these entropy perturbation variables should be used to match observations, in particular for the subtle case of non-linear sigma models. We first reviewed all the ways to describe non-adiabatic modes: ($i$) the total entropy perturbation $S$ which can be decomposed in an intrinsic contribution $S_{\rm int}$ and a relative one $S_{\rm rel}$; ($ii$) the relative entropy perturbation between two fluids $S_{ij}$ defined as the difference between the corresponding curvature perturbations, $S_{ij} = \zeta_i-\zeta_j$; ($iii$) the fluctuation $\delta s$ of the entropy field corresponding to fluctuations orthogonal to the background inflationary trajectory in field space. This last variable, despite being very useful to perform an efficient numerical integration of the field equations during inflation, is not the right one to match data which are instead expressed in terms of $S_{ij}$. Similar considerations apply to the total entropy perturbation $S$ since it is proportional to $\delta s$. 

We therefore argued that the right variable to be studied is $S_{ij}$ which should be evolved from the inflationary epoch to the post-inflationary era until CMB decoupling where observations are performed. This process clearly requires a detailed understanding of the reheating epoch and how the fluctuations of the microscopic degrees of freedom get transferred to density and pressure perturbations of the different fluids (photons, baryons, cold dark matter and neutrinos) which characterise the post-inflationary history of our Universe. 

We noticed that the use of $S$ or $\delta s$ instead of $S_{ij}$ in general can yield results which are wrong even by several orders of magnitude. Moreover, in the case of kinetically coupled light spectator fields which can arise naturally in string inflation, the mass of $\delta s$ fluctuations can become tachyonic, signalling the emergence of a potentially dangerous destabilisation effect due to the growth of these isocurvature modes. In this situation, the relative entropy perturbation $S_{ij}$ becomes also pathological since it blows up due to the fact that during inflation the energy density of the spectator fields very quickly goes to zero. 

In a simple 2-field model with a Starobinsky-like inflationary potential and a light spectator field, we showed that the growth of these non-adiabatic modes is unphysical and the associated destabilisation effect is spurious since it is caused just by the use of entropy variables which in this case become ill-defined. In fact, $\delta s$ increases after horizon-exit just because the vector normal to the background trajectory diverges, while no anomalous growth of isocurvature fluctuations is seen when considering scalar fluctuations in the original field basis. We therefore defined a so-called improved entropy perturbation $\hat{S}_{ij}$ associated to the relative entropy perturbation $S_{\rm rel}$ which remains always well-behaved and decays during inflation, showing that the dynamics of this class of models becomes effectively single field, with no production of isocurvature fluctuations. In order to perform a detailed matching with observations, one should therefore study the evolution of $\hat{S}_{ij}$ during inflation and the subsequent reheating epoch, translating at the end its value into the standard relative entropy perturbation $S_{n\gamma}$ between photons and a generic $n$ species since this quantity is now finite and it is the one used to express the isocurvature fraction $\beta_{\rm iso}$ constrained by data.

Let us finally stress that all entropy variables described in this paper are gauge invariant, even in the generic case of a curved field space. Hence in principle they are all good candidates to describe physical quantities. However well-defined physical quantities should be not just gauge invariant but also finite. The failure to satisfy this last requirement is the reason why $\delta s$ and $S_{ij}$ are not the proper entropy variables to be used to confront observations, as emerges clearly in the case of light kinetically coupled spectator fields which is characterised by unphysical instabilities. As already pointed out, the spurious nature of this apparent destabilisation effect manifests itself in the basis-dependence of the time evolution of isocurvature perturbations which diverge in the kinematic basis whereas decay in the field space one. A more sophisticated analysis of the proper physical variable that should be used to describe entropy perturbations in multifield models should therefore be basis-independent, i.e. invariant under field reparametrisations. This investigation is however beyond the scope of our paper, and so we leave it for future work.

\section*{Acknowledgements}

We would like to thank Katy Clough, Evangelos Sfakianakis and Yvette Welling for useful discussions. FM is funded by a UKRI/EPSRC Stephen Hawking fellowship, grant reference EP/T017279/1 and partially supported by the STFC consolidated grant ST/P000681/1.

\appendix

\section{Perturbation theory and gauge invariance in curved field space}
\label{AppA}

\subsection{General framework in spatially flat gauge}

Let us start by considering the most generic perturbed line element:
\be
ds^2=-(1+2\Phi)dt^2+2aB_i dt dx^i + a^2\left[(1-2\Psi)\delta_{ij} + E_{ij}\right]dx^idx^j\,.
\label{dsgen}
\ee
Combining metric perturbations with scalar field perturbations:
\be 
\varphi^A = \phi^A+\delta\phi^A\,,
\ee
we can compute the perturbed Einstein equations as:
\be
\delta G_\mu^\nu=\delta T_\mu^\nu\,.
\label{eq:pertEinstein}
\ee
Considering a curved field space, we list the resulting equations below. The $(0,0)$ component of (\ref{eq:pertEinstein}) gives:
\bea
&&\left(6H\partial_t -2\frac{\partial_{kk}}{a^2}\right)\Psi +2\frac{H}{a} \partial_{kk}B +\left(6H^2-\dot{\phi}_A\dot{\phi}^A\right)\Phi -\frac{1}{2a^2}\partial_{ki}E_{ki} \nonumber \\
&&+ V_A \delta \phi^A+\dot{\phi}^A G_{AB} \dot{\delta \phi}^B+\frac{1}{2}\dot{\phi}^A\dot{\phi}^BG_{AB,C}\delta\phi^C=0\,.
\eea
From the $(i,0)$ component we get:
\be
2\partial_i \dot{\Psi} +2 H \partial_i\Phi+\frac{1}{2}\partial_j \dot{E}_{ij}  -\dot{\phi}^A G_{AB} \partial_i \delta\phi^B=0\,.
\ee
The spatial $(i,j)$ components with $i\neq j$ give rise to:
\be
\partial_i\partial_j \Psi-\partial_i\partial_j \Phi+a^2\left(\frac{\partial_{tt}}{2}+\frac{3}{2}H\partial_t-\frac{\partial_{kk}}{2a^2}\right) E_{ij}+\frac12\left(\partial_{ik}E_{jk}+\partial_{jk}E_{ik}\right)-a\left(\partial_t+2H\right)\partial_{ij}B=0 \,,
\ee
while if $i=j$ we find:
\bea
&&\left(6\partial_{tt} -2 \frac{\partial_{ii}}{a^2} + 18H\partial_t \right)\Psi+\frac{2}{a}\left(\partial_t +2H\right) \partial_{kk}B+ 
\left(2 \frac{\partial_{ii}}{a^2} +6 H\partial_t +6H^2 + 12\frac{\ddot{a}}{a} + 3 \dot{\phi}^A\dot{\phi}_A\right)\Phi \nonumber \\
&&-\frac{1}{2}\frac{\partial_{ki}E_{ik}}{a^2}+ 3\left(V_A\delta\phi^A -\dot{\delta\phi}^A\dot{\phi}^B G_{AB}- \dot{\phi}^A\dot{\phi}^B G_{AB,C}\delta\phi^C\right)=0\,.
\eea
In what follows we use the spatially flat gauge, $E=\Psi=0$, that is a very convenient setup for computing inflationary perturbations. After two spatial integrations and fixing arbitrary integration functions of time to zero, the component $(i,j)$ with $i\neq j$ reduces to:
\be
\Phi+2 \dot{a} B+ a \dot{B}=0\,.
\ee
After one spatial integration, the $(0,i)$ component becomes:
\be
H \Phi =\frac12 \dot{\phi}_A\delta\phi^A\,.
\label{eq:Phieom}
\ee
Finally the $(0,0)$ component is given by:
\be
6\Phi \left(\frac{\dot{a}}{a}\right)^2+2\frac{H}{a} B_{ii}=\Phi \dot{\phi}_A \dot{\phi}^A - \dot{\phi}_A \delta\dot\phi^A -\frac12 \dot{\phi}^A \dot{\phi}^B \partial_C G_{AB} \delta \phi^C - V_A \delta\phi^A\,,
\ee
and, making use of the previous equations, becomes:
\be
\partial_i\partial_i B=\frac{a}{2H}\left[\left(\frac{\dot\phi_0^2 }{2H}\dot{\phi}_A -3H\dot{\phi}_A - \frac12 G_{BC,A}\dot{\phi}^B \dot{\phi}^C - V_A \right)\delta\phi^A-\dot{\phi}_A \delta\dot\phi^A\right].
\ee
Working in the spatially flat gauge, the gauge invariant MS variables coincide with field perturbations:
\be
Q^A \equiv \delta \phi^A + \frac{\dot{\phi}^A}{H}\Psi \equiv \delta \phi^A\,,
\ee
and the Klein-Gordon equation for perturbations is given by:
\bea
&& D_tD_t Q^B -a^{-2}\partial_i\partial_iQ^B + 3HD_t Q^B +\left[R_{C A D}^B \dot\phi^C \dot\phi^D +G^{B L}V_{;LA} \right. \nonumber \\
&& \left. +\frac{1}{H}\left(\dot{\phi}_A V^B+V_A \dot{\phi}^B \right)+\dot{\phi}_A \dot{\phi}^B \left(3-\frac{\dot\phi_0^2}{2H^2}\right) \right]Q^A =0\,,
\label{eq:Qequation}
\eea
where $D_t Q^A=\dot Q^A + \Gamma^A_{BC} Q^B \dot{\phi}^C$, $V_{;L A}=V_{L A}-\Gamma_{L A}^B V_B$ are covariant derivatives and $R_{C A D}^B$ is the Riemann tensor of the field space. This is the well-known gauge invariant equation for field perturbations \cite{Sasaki:1995aw}.

\subsection{Gauge invariance of entropy variables}
\label{app:gaugeinvariance}

In this appendix we prove the gauge invariance, at first order in cosmological perturbation theory, of all different entropy variables mentioned in the main text. We shall consider an $N$-dimensional non-linear sigma model described by the Lagrangian (\ref{eq:L}) in the simple case where the metric is diagonal, i.e. $G_{ij}(\phi^k)=G_{i}(\phi^k) \delta_{ij}$, with a generic potential which we arbitrarily decompose as $V(\phi^k)=\sum_i V^{(i)}(\phi^k)$. Similarly to the spacetime metric, we split the fields in background and fluctuation components as $\phi^i(x^\mu)=\phi^i(x^0)+\delta\phi^i(x^\mu)$.

The background evolution in a FLRW spacetime is given by:
\be
G_{(i)} \ddot {\phi^i}+3H G_{(i)} \dot{\phi^i}+ G_{{(i)},j} \dot{\phi^i} \dot{\phi^j}-\frac12 G_{j,i} \dot{\phi^j}^2+ V_i=0 \,,
\label{EOM}
\ee
where we adopt the convention that contracted (upper and lower) indices are summed, unless an index is written inside round brackets, i.e. contraction of $i$ with $(i)$ implies no sum. Let us define the background density and pressure components of the system as:
\be
\rho_i=\frac12 G_{(i)} \dot{\phi^i}^2+V^{(i)} \qquad\text{and}\qquad
P_i =\frac12 G_{(i)} \dot{\phi^i}^2-V^{(i)} \,,
\ee
which sum up to the total quantities $\rho=\sum_i \rho_i$ and $P=\sum_i P_i$. Using the equations of motion~\eqref{EOM} one finds:
\bea
\dot \rho_i&=& \frac12 G_{(i),j} \dot {\phi^j} \dot{\phi^i}^2 +G_{(i)} \dot {\phi^i} \ddot {\phi^i} + V^{(i)}_j \dot {\phi^j}\nonumber\\
&=& -3H G_{(i)} \dot {\phi^i}^2-\frac12 G_{(i),j} \dot {\phi^j} \dot{\phi^i}^2+\frac12 G_{j,(i)} \dot {\phi^j}^2 \dot{\phi^i}
+V^{(i)}_j \dot {\phi^j}- V_i \dot\phi^{(i)}\,,
\eea
which trivially verifies the background relation:
\be
\dot \rho = \sum_i \dot \rho_i = -3 H G_i \dot {\phi^i}^2 = -3H \left(\rho+P\right),
\ee
and:
\be
\dot P_i= \dot \rho_i - 2 V^{(i)}_j \dot {\phi^j}\,.
\ee
The density and pressure fluctuations can instead be written in terms of the scalar and spacetime metric fluctuations as:
\bea
\delta \rho_i &=& -\Phi G_{(i)} \dot {\phi^i}^2 +  \frac12 G_{(i),j} \dot {\phi^i}^2 \delta \phi^j+G_{(i)} \dot {\phi^i} \delta \dot{\phi^i}+ V^{(i)}_j \delta \phi^j\,, \\
\delta P_i &=& \delta\rho_i- 2 V^{(i)}_j \delta \phi^j\,.
\eea

Let us now consider a gauge transformation induced by a change of reference frame $x^\mu \to x^\mu+\epsilon^\mu$. At linear order in the fluctuations $\epsilon^\mu$ and $\delta\phi^i$ we obtain:
\be
\phi^i \to \phi^i\,,\qquad
\delta \phi^i \to \delta \phi^i - \dot {\phi^i} \epsilon^0\,, \qquad \Phi \to \Phi - \dot \epsilon^0\,, \qquad \Psi \to \Psi + H \epsilon^0 \,,
\label{gaugevar}
\ee
where $\Phi$ and $\Psi$ are the two metric scalar fluctuations introduced in (\ref{dsgen}). The induced change in the density fluctuations goes as follows:
\bea
\delta \rho_i &\to& -(\Phi - \dot \epsilon^0) G_{(i)} \dot{ \phi^i}^2 +  \frac12 G_{(i),j} \dot {\phi^i}^2 (\delta \phi^j -\dot {\phi^j} \epsilon^0)+
G_{(i)} \dot {\phi^i} (\delta \dot{\phi^i} -\dot {\phi^i} \dot \epsilon^0 - \ddot {\phi^i} \epsilon^0)+ V^{(i)}_j (\delta \phi^j - \dot {\phi^j} \epsilon^0) \nonumber \\
&=& \delta \rho_i - \epsilon^0 \dot{\rho}_i\,,
\label{varrho}
\eea
where we used again the equations of motion~\eqref{EOM}. Similarly, it can be easily seen that the pressure fluctuations transform at linear order as:
\be
\delta P_i\to  \delta P_i - \epsilon^0 \dot{P}_i\,.
\label{varP}
\ee

With these relations at hands, it is now straightforward to show the gauge invariance (at first order) of all different entropy variables introduced in Sec. \ref{sec:EntropyPerturbations}.

\subsubsection*{Total entropy}

The total entropy perturbation (\ref{eq:S}) can also be expressed as
\be
S=H \left( \frac{\delta P}{\dot P} -\frac{\delta \rho}{\dot \rho} \right) \,,
\ee
and using (\ref{varrho}) and (\ref{varP}) we find:
\be
S\to H \left( \frac{\delta P}{\dot P} -\epsilon^0 -\frac{\delta \rho}{\dot \rho} +\epsilon^0 \right) =S\,.
\ee

\subsubsection*{Intrinsic entropy}

The intrinsic entropy perturbation for a given fluid $i$ given by (\ref{eq:Sint}) is also clearly gauge invariant since it can be rewritten as:
\be
S_{{\rm int},i}=H \frac{\dot P_i}{\dot P} \left( \frac{\delta P_i}{\dot P_i} -\frac{\delta \rho_i}{\dot \rho_i} \right) \to 
H \frac{\dot P_i}{\dot P} \left( \frac{\delta P_i}{\dot P_i} -\epsilon^0 -\frac{\delta \rho_i}{\dot \rho_i} +\epsilon^0\right) = S_{{\rm int},i}\,.
\ee
Thus also the total intrinsic entropy perturbation $S_{\rm int}=\sum_i S_{{\rm int},i}$ of a system with several components turns out to be gauge invariant.

\subsubsection*{Relative entropy}

Given that the relative entropy perturbation $S_{\rm rel}$ is the difference between two gauge invariant quantities, $S$ and $S_{\rm int}$, it definitely turns out to be gauge invariant. In particular, each of the $\frac{N(N-1)}{2}$ relative entropy perturbations $S_{ij}$ introduced in (\ref{eq:Sab}) is gauge invariant since it is defined as $S_{ij}= \zeta_i-\zeta_j$ in terms of $N$ distinct gauge invariant curvature perturbations (using (\ref{gaugevar}) and (\ref{varrho})):
\be
\zeta_i = - \Psi-H \,\frac{\delta \rho_i}{\dot \rho_i}\to - \Psi - H\epsilon^0 -H \,\frac{\delta \rho_i}{\dot \rho_i} + H\epsilon^0 = \zeta_i\,.
\ee

\subsubsection*{Entropy field}

The fluctuation of the entropy field $\delta s=N^i\delta\phi_i$ is defined in (\ref{deltas}) in terms of the vector normal to the background trajectory $N^i= D_t T^i / || D_t \vec{T}||$ which, according to (\ref{eq:TN}), can in turn be derived from the covariant derivative of the tangent vector given by: 
\be
D_t T^i=\frac{1}{\dot{\phi}_0^3 G_i} \left(V_i \dot{\phi}_0^2 - G_{(i)} \dot{\phi^i} V_j \dot{\phi^j} \right)\,.
\ee
From these expressions it can be easily checked that $G_{ij} T^i N^j =0$, as expected from the orthogonality condition. Focusing for simplicity on the 2-field case we find:
\be
D_t \vec{T} = \begin{pmatrix}
\dot{\phi^2}\,G_1^{-1}\\
 -  \dot{\phi^1}\,G_2^{-1} \\
\end{pmatrix}
 \frac{1}{\dot{\phi}_0^3}\,\left(V_1 G_2 \dot{\phi^2} - V_2 G_1 \dot{\phi^1}\right)
\ee
and:
\be
\vec{N}= \frac{\sqrt{G_1 G_2}}{\dot{\phi}_0}
\begin{pmatrix}
\dot{\phi^2}\,G_1^{-1}\\
 -  \dot{\phi^1}\,G_2^{-1} \\
\end{pmatrix}\,.
\ee
It is now easy to obtain the following expression for $\delta s$:
\be
\delta s = G_{ij} N^i \delta \phi^j= G_1 N^1 \delta \phi^1 +G_2 N^2 \delta \phi^2= \sqrt{G_1 G_2} \left(\frac{\dot{\phi^1} \dot{\phi^2}}{\dot\phi_0}\right) \left( \frac{\delta \phi^1}{\dot{\phi^1}} - \frac{\delta \phi^2}{\dot{\phi^2}} \right) \,,
\ee
which using (\ref{gaugevar}) shows clearly that this entropy variable is gauge invariant:
\be
\delta s \to  
\sqrt{G_1 G_2} \left(\frac{\dot{\phi^1} \dot{\phi^2}}{\dot\phi_0}\right) \left( \frac{\delta \phi^1}{\dot{\phi^1}}-\epsilon^0 - \frac{\delta \phi^2}{\dot{\phi^2}} +\epsilon^0\right) =\delta s \,.
\ee

\section{Single field limit of density perturbations}
\label{sec:singlefield}

In single field inflation, the dynamics of the scalar perturbations is determined by the (Fourier space) MS equation:
\be
u''+\left(k^2-\frac{z''}{z}\right) u=0\,,
\ee
where the canonically normalised scalar perturbation is defined as $u\equiv a Q$, primes denote derivatives with respect to conformal time, and $z\equiv \sqrt{2\epsilon} a$. In order to make contact with the multifield results described in the main text, it is useful to rewrite the single field MS equation using cosmic time and $Q$ instead of $u$: 
\be
\ddot{Q}+3H \dot{Q}+\left ( \frac{k^2}{a^2}-\frac{z''}{z a^2}+H^2+\frac{\ddot{a}}{a}\right )Q=0\,,
\ee
prompting the definition:
\be
m_Q^2\equiv -\frac{z''}{z a^2}+H^2+\frac{\ddot{a}}{a}\,,
\ee
which using the exact relation:
\be
\frac{z''}{z}=(aH)^2 \left(2-\epsilon+\frac{3 \eta}{2}-\frac{\epsilon \eta}{2} +\frac{\eta^2}{4}+\frac{\eta \kappa}{2}\right),
\ee
and the definitions of the Hubble slow-roll parameters, can be recast as:
\be
\frac{m_Q^2}{H^2}=-\frac{3 \eta}{2}+\frac{\epsilon \eta}{2}-\frac{\eta^2}{4}-\frac{\eta \kappa}{2}\,.
\label{eq:mQ2}
\ee
In order to relate ${(\mathbb{M}^2)^1}_1$ to the above result it is necessary to translate between the potential slow-roll parameters:
\be
\epsilon_V\equiv \frac{1}{2} \left(\frac{V'}{V}\right)^2 \qquad\text{and}\qquad \eta_V\equiv\frac{V''}{V}\,,
\ee
and the Hubble slow-roll parameters. Using the background equations of motion one may show that:
\be
\epsilon=\epsilon_V \,\frac{3-\epsilon}{3-\epsilon +\frac{\eta}{2}}\,.
\label{eq:epsV}
\ee
Taking the time derivative of the background Klein-Gordon equation and using the definitions of the potential and Hubble slow-roll parameters one also can show that:
\be
\eta_V=\frac{3(2 \epsilon-\frac{\eta}{2})-2\epsilon^2+\frac{5}{2} \epsilon \eta -\frac{\eta}{4}(\eta+2\kappa) }{3-\epsilon}\,.
\label{eq:etaV}
\ee
Notice that these results are exact and involve no slow-roll expansion, depending only on the definitions of the various slow-roll parameters. Using these results we can show that \eqref{eq:mQ22Field} reduces to \eqref{eq:mQ2}:
\be
\begin{split}
{\left(\mathbb{M}^2\right)^1}_1&= V_{11}+4\frac{\epsilon H }{\dot\phi_0}\,V_1+2\epsilon(3-\epsilon) H^2\\
&=V \eta_V-4 \epsilon H^3\left(3-\epsilon+\frac{\eta}{2}\right)+2\epsilon(3-\epsilon) H^2\\
&=H^2\left(-\frac{3 \eta}{2}+\frac{\epsilon \eta}{2}-\frac{\eta^2}{4}-\frac{\kappa \eta}{2}\right),
\end{split}
\ee
where we have eliminated derivatives of $V$ in favour of the corresponding slow-roll parameters and in the first and second steps we have used \eqref{eq:epsV} and \eqref{eq:etaV} respectively.


\begin{thebibliography}{10}

\bibitem{Guth:1980zm}
A.~H.~Guth,
``The Inflationary Universe: A Possible Solution to the Horizon and Flatness Problems,''
Phys. Rev. D \textbf{23} (1981), 347-356
doi:10.1103/PhysRevD.23.347

\bibitem{Linde:1981mu}
A.~D.~Linde,
``A New Inflationary Universe Scenario: A Possible Solution of the Horizon, Flatness, Homogeneity, Isotropy and Primordial Monopole Problems,''
Phys. Lett. B \textbf{108} (1982), 389-393
doi:10.1016/0370-2693(82)91219-9

\bibitem{Albrecht:1982wi}
A.~Albrecht and P.~J.~Steinhardt,
``Cosmology for Grand Unified Theories with Radiatively Induced Symmetry Breaking,''
Phys. Rev. Lett. \textbf{48} (1982), 1220-1223
doi:10.1103/PhysRevLett.48.1220

\bibitem{Baumann:2009ds}
D.~Baumann,
``Inflation,''
doi:10.1142/9789814327183\_0010
[arXiv:0907.5424 [hep-th]].

\bibitem{Akrami:2018odb}
Y.~Akrami \textit{et al.} [Planck],
``Planck 2018 results. X. Constraints on inflation,''
Astron. Astrophys. \textbf{641} (2020), A10
doi:10.1051/0004-6361/201833887
[arXiv:1807.06211 [astro-ph.CO]].

\bibitem{Linde:1983gd}
A.~D.~Linde,
``Chaotic Inflation,''
Phys. Lett. B \textbf{129} (1983), 177-181
doi:10.1016/0370-2693(83)90837-7

\bibitem{Kaiser:2010ps}
D.~I.~Kaiser,
``Conformal Transformations with Multiple Scalar Fields,''
Phys. Rev. D \textbf{81} (2010), 084044
doi:10.1103/PhysRevD.81.084044
[arXiv:1003.1159 [gr-qc]].

\bibitem{Kaiser:2012ak}
D.~I.~Kaiser, E.~A.~Mazenc and E.~I.~Sfakianakis,
``Primordial Bispectrum from Multifield Inflation with Nonminimal Couplings,''
Phys. Rev. D \textbf{87} (2013), 064004
doi:10.1103/PhysRevD.87.064004
[arXiv:1210.7487 [astro-ph.CO]].

\bibitem{Schutz:2013fua}
K.~Schutz, E.~I.~Sfakianakis and D.~I.~Kaiser,
``Multifield Inflation after Planck: Isocurvature Modes from Nonminimal Couplings,''
Phys. Rev. D \textbf{89} (2014) no.6, 064044
doi:10.1103/PhysRevD.89.064044
[arXiv:1310.8285 [astro-ph.CO]].

\bibitem{DiMarco:2002eb}
F.~Di Marco, F.~Finelli and R.~Brandenberger,
``Adiabatic and isocurvature perturbations for multifield generalized Einstein models,''
Phys. Rev. D \textbf{67} (2003), 063512
doi:10.1103/PhysRevD.67.063512
[arXiv:astro-ph/0211276 [astro-ph]].

\bibitem{Svrcek:2006yi}
P.~Svrcek and E.~Witten,
``Axions In String Theory,''
JHEP \textbf{06} (2006), 051
doi:10.1088/1126-6708/2006/06/051
[arXiv:hep-th/0605206 [hep-th]].

\bibitem{Conlon:2006tq}
J.~P.~Conlon,
``The QCD axion and moduli stabilisation,''
JHEP \textbf{05} (2006), 078
doi:10.1088/1126-6708/2006/05/078
[arXiv:hep-th/0602233 [hep-th]].

\bibitem{Arvanitaki:2009fg}
A.~Arvanitaki, S.~Dimopoulos, S.~Dubovsky, N.~Kaloper and J.~March-Russell,
``String Axiverse,''
Phys. Rev. D \textbf{81} (2010), 123530
doi:10.1103/PhysRevD.81.123530
[arXiv:0905.4720 [hep-th]].

\bibitem{Cicoli:2012sz}
M.~Cicoli, M.~Goodsell and A.~Ringwald,
``The type IIB string axiverse and its low-energy phenomenology,''
JHEP \textbf{10} (2012), 146
doi:10.1007/JHEP10(2012)146
[arXiv:1206.0819 [hep-th]].

\bibitem{Polarski:1994rz}
D.~Polarski and A.~A.~Starobinsky,
``Isocurvature perturbations in multiple inflationary models,''
Phys. Rev. D \textbf{50} (1994), 6123-6129
doi:10.1103/PhysRevD.50.6123
[arXiv:astro-ph/9404061 [astro-ph]].

\bibitem{Langlois:1999dw}
D.~Langlois,
``Correlated adiabatic and isocurvature perturbations from double inflation,''
Phys. Rev. D \textbf{59} (1999), 123512
doi:10.1103/PhysRevD.59.123512
[arXiv:astro-ph/9906080 [astro-ph]].

\bibitem{Martin:2021frd}
J.~Martin and L.~Pinol,
``Opening the reheating box in multifield inflation,''
[arXiv:2105.03301 [astro-ph.CO]].

\bibitem{Wands:2000dp}
D.~Wands, K.~A.~Malik, D.~H.~Lyth and A.~R.~Liddle,
``A New approach to the evolution of cosmological perturbations on large scales,''
Phys. Rev. D \textbf{62} (2000), 043527
doi:10.1103/PhysRevD.62.043527
[arXiv:astro-ph/0003278 [astro-ph]].

\bibitem{Cicoli:2018ccr}
M.~Cicoli, V.~Guidetti, F.~G.~Pedro and G.~P.~Vacca,
``A geometrical instability for ultra-light fields during inflation?,''
JCAP \textbf{12} (2018), 037
doi:10.1088/1475-7516/2018/12/037
[arXiv:1807.03818 [hep-th]].

\bibitem{Cicoli:2019ulk}
M.~Cicoli, V.~Guidetti and F.~G.~Pedro,
``Geometrical Destabilisation of Ultra-Light Axions in String Inflation,''
JCAP \textbf{05} (2019), 046
doi:10.1088/1475-7516/2019/05/046
[arXiv:1903.01497 [hep-th]].

\bibitem{Renaux-Petel:2015mga}
S.~Renaux-Petel and K.~Turzy\'nski,
``Geometrical Destabilization of Inflation,''
Phys. Rev. Lett. \textbf{117} (2016) no.14, 141301
doi:10.1103/PhysRevLett.117.141301
[arXiv:1510.01281 [astro-ph.CO]].

\bibitem{Huston:2011fr}
I.~Huston and A.~J.~Christopherson,
``Calculating Non-adiabatic Pressure Perturbations during Multi-field Inflation,''
Phys. Rev. D \textbf{85} (2012), 063507
doi:10.1103/PhysRevD.85.063507
[arXiv:1111.6919 [astro-ph.CO]].

\bibitem{Huston:2013kgl}
I.~Huston and A.~J.~Christopherson,
``Isocurvature Perturbations and Reheating in Multi-Field Inflation,''
[arXiv:1302.4298 [astro-ph.CO]].

\bibitem{Gordon:2000hv}
C.~Gordon, D.~Wands, B.~A.~Bassett and R.~Maartens,
``Adiabatic and entropy perturbations from inflation,''
Phys. Rev. D \textbf{63} (2000), 023506
doi:10.1103/PhysRevD.63.023506
[arXiv:astro-ph/0009131 [astro-ph]].

\bibitem{Achucarro:2010da}
A.~Achucarro, J.~O.~Gong, S.~Hardeman, G.~A.~Palma and S.~P.~Patil,
``Features of heavy physics in the CMB power spectrum,''
JCAP \textbf{01} (2011), 030
doi:10.1088/1475-7516/2011/01/030
[arXiv:1010.3693 [hep-ph]].

\bibitem{Cremonini:2010sv}
S.~Cremonini, Z.~Lalak and K.~Turzynski,
``On Non-Canonical Kinetic Terms and the Tilt of the Power Spectrum,''
Phys. Rev. D \textbf{82} (2010), 047301
doi:10.1103/PhysRevD.82.047301
[arXiv:1005.4347 [hep-th]].

\bibitem{Wands:2002bn}
D.~Wands, N.~Bartolo, S.~Matarrese and A.~Riotto,
``An Observational test of two-field inflation,''
Phys. Rev. D \textbf{66} (2002), 043520
doi:10.1103/PhysRevD.66.043520
[arXiv:astro-ph/0205253 [astro-ph]].

\bibitem{Gong:2011uw}
J.~O.~Gong and T.~Tanaka,
``A covariant approach to general field space metric in multi-field inflation,''
JCAP \textbf{03} (2011), 015
[erratum: JCAP \textbf{02} (2012), E01]
doi:10.1088/1475-7516/2012/02/E01
[arXiv:1101.4809 [astro-ph.CO]].

\bibitem{Christodoulidis:2019mkj}
P.~Christodoulidis, D.~Roest and E.~I.~Sfakianakis,
``Attractors, Bifurcations and Curvature in Multi-field Inflation,''
JCAP \textbf{08} (2020), 006
doi:10.1088/1475-7516/2020/08/006
[arXiv:1903.03513 [gr-qc]].

\bibitem{Achucarro:2016fby}
A.~Ach\'ucarro, V.~Atal, C.~Germani and G.~A.~Palma,
``Cumulative effects in inflation with ultra-light entropy modes,''
JCAP \textbf{02} (2017), 013
doi:10.1088/1475-7516/2017/02/013
[arXiv:1607.08609 [astro-ph.CO]].

\bibitem{Conlon:2005jm}
J.~P.~Conlon and F.~Quevedo,
``Kahler moduli inflation,''
JHEP \textbf{01} (2006), 146
doi:10.1088/1126-6708/2006/01/146
[arXiv:hep-th/0509012 [hep-th]].

\bibitem{Cicoli:2008gp}
M.~Cicoli, C.~P.~Burgess and F.~Quevedo,
``Fibre Inflation: Observable Gravity Waves from IIB String Compactifications,''
JCAP \textbf{03} (2009), 013
doi:10.1088/1475-7516/2009/03/013
[arXiv:0808.0691 [hep-th]].

\bibitem{Burgess:2010bz}
C.~P.~Burgess, M.~Cicoli, M.~Gomez-Reino, F.~Quevedo, G.~Tasinato and I.~Zavala,
``Non-standard primordial fluctuations and nongaussianity in string inflation,''
JHEP \textbf{08} (2010), 045
doi:10.1007/JHEP08(2010)045
[arXiv:1005.4840 [hep-th]].

\bibitem{Cicoli:2011ct}
M.~Cicoli, F.~G.~Pedro and G.~Tasinato,
``Poly-instanton Inflation,''
JCAP \textbf{12} (2011), 022
doi:10.1088/1475-7516/2011/12/022
[arXiv:1110.6182 [hep-th]].

\bibitem{Cicoli:2012cy}
M.~Cicoli, G.~Tasinato, I.~Zavala, C.~P.~Burgess and F.~Quevedo,
``Modulated Reheating and Large Non-Gaussianity in String Cosmology,''
JCAP \textbf{05} (2012), 039
doi:10.1088/1475-7516/2012/05/039
[arXiv:1202.4580 [hep-th]].

\bibitem{Cicoli:2016chb}
M.~Cicoli, D.~Ciupke, S.~de Alwis and F.~Muia,
``$\alpha'$ Inflation: moduli stabilisation and observable tensors from higher derivatives,''
JHEP \textbf{09} (2016), 026
doi:10.1007/JHEP09(2016)026
[arXiv:1607.01395 [hep-th]].

\bibitem{Cicoli:2016xae}
M.~Cicoli, F.~Muia and P.~Shukla,
``Global Embedding of Fibre Inflation Models,''
JHEP \textbf{11} (2016), 182
doi:10.1007/JHEP11(2016)182
[arXiv:1611.04612 [hep-th]].

\bibitem{Cicoli:2017axo}
M.~Cicoli, D.~Ciupke, V.~A.~Diaz, V.~Guidetti, F.~Muia and P.~Shukla,
``Chiral Global Embedding of Fibre Inflation Models,''
JHEP \textbf{11} (2017), 207
doi:10.1007/JHEP11(2017)207
[arXiv:1709.01518 [hep-th]].

\bibitem{Cicoli:2020bao}
M.~Cicoli and E.~Di Valentino,
``Fitting string inflation to real cosmological data: The fiber inflation case,''
Phys. Rev. D \textbf{102} (2020) no.4, 043521
doi:10.1103/PhysRevD.102.043521
[arXiv:2004.01210 [astro-ph.CO]].

\bibitem{Malik:2004tf}
K.~A.~Malik and D.~Wands,
``Adiabatic and entropy perturbations with interacting fluids and fields,''
JCAP \textbf{02} (2005), 007
doi:10.1088/1475-7516/2005/02/007
[arXiv:astro-ph/0411703 [astro-ph]].

\bibitem{Malik:2002jb}
K.~A.~Malik, D.~Wands and C.~Ungarelli,
``Large scale curvature and entropy perturbations for multiple interacting fluids,''
Phys. Rev. D \textbf{67} (2003), 063516
doi:10.1103/PhysRevD.67.063516
[arXiv:astro-ph/0211602 [astro-ph]].

\bibitem{Price:2014xpa}
L.~C.~Price, J.~Frazer, J.~Xu, H.~V.~Peiris and R.~Easther,
``MultiModeCode: An efficient numerical solver for multifield inflation,''
JCAP \textbf{03} (2015), 005
doi:10.1088/1475-7516/2015/03/005
[arXiv:1410.0685 [astro-ph.CO]].

\bibitem{Hwang:2000jh}
J.~c.~Hwang and H.~Noh,
``Cosmological perturbations with multiple scalar fields,''
Phys. Lett. B \textbf{495} (2000), 277-283
doi:10.1016/S0370-2693(00)01253-3
[arXiv:astro-ph/0009268 [astro-ph]].

\bibitem{Montandon:2020kuk}
T.~Montandon, G.~Patanchon and B.~van Tent,
``Isocurvature modes: joint analysis of the CMB power spectrum and bispectrum,''
JCAP \textbf{01} (2021), 004
doi:10.1088/1475-7516/2021/01/004
[arXiv:2007.05457 [astro-ph.CO]].

\bibitem{Bartolo:2001rt}
N.~Bartolo, S.~Matarrese and A.~Riotto,
``Adiabatic and isocurvature perturbations from inflation: Power spectra and consistency relations,''
Phys. Rev. D \textbf{64} (2001), 123504
doi:10.1103/PhysRevD.64.123504
[arXiv:astro-ph/0107502 [astro-ph]].

\bibitem{Sasaki:1995aw}
M.~Sasaki and E.~D.~Stewart,
``A General analytic formula for the spectral index of the density perturbations produced during inflation,''
Prog. Theor. Phys. \textbf{95} (1996), 71-78
doi:10.1143/PTP.95.71
[arXiv:astro-ph/9507001 [astro-ph]].

\bibitem{Starobinsky:1980te}
A.~A.~Starobinsky,
``A New Type of Isotropic Cosmological Models Without Singularity,''
Phys. Lett. B \textbf{91} (1980), 99-102
doi:10.1016/0370-2693(80)90670-X

\bibitem{Cicoli:2020cfj}
M.~Cicoli, G.~Dibitetto and F.~G.~Pedro,
``New accelerating solutions in late-time cosmology,''
Phys. Rev. D \textbf{101} (2020) no.10, 103524
doi:10.1103/PhysRevD.101.103524
[arXiv:2002.02695 [gr-qc]].

\bibitem{Cicoli:2020noz}
M.~Cicoli, G.~Dibitetto and F.~G.~Pedro,
``Out of the Swampland with Multifield Quintessence?,''
JHEP \textbf{10} (2020), 035
doi:10.1007/JHEP10(2020)035
[arXiv:2007.11011 [hep-th]].
  
\end{thebibliography}
\end{document}